\documentclass[aps,prl,reprint,floatfix,showpacs]{revtex4-1}
\usepackage{graphicx}

\newcommand*{\pr}[1]{\mathcal{#1}}
\newcommand*{\refneq}[1]{(\ref{#1})}
\newcommand*{\refneqs}[1]{(\ref{#1})}
\newcommand*{\refeq}[1]{Eq.\ (\ref{#1})}
\newcommand*{\reflet}[1]{Eq.\ (#1)}
\newcommand*{\refsup}[1]{\cite{supplement}}
\newcommand*{\Psilamnl}{1}
\newcommand*{\SpinFun}{2}
\newcommand*{\Wk}{3}
\newcommand*{\Fkexp}{8}
\newcommand*{\hatrhok}{6}
\newcommand*{\hatgk}{7}
\newcommand*{\rhotilklam}{9}
\newcommand*{\Eavzero}{10}
\newcommand*{\bargkzero}{11}

\begin{document}


\title{Permutation symmetry in spinor quantum gases: selection rules, conservation laws, and correlations}


\author{Vladimir A. Yurovsky}
\affiliation{School of Chemistry, Tel Aviv University, 69978 Tel Aviv, Israel}
\affiliation{Institute for Theoretical Physics, University of California, Santa Barbara, CA 93106 USA}

\date{\today}

\begin{abstract}
Many-body systems of identical arbitrary-spin particles, with separable spin and spatial degrees of freedom, are considered. Their eigenstates can be classified by Young diagrams, corresponding to non-trivial permutation symmetries (beyond the conventional paradigm of symmetric--antisymmetric states). 

The present work obtains (a) selection rules for additional non-separable (dependent on spins and coordinates) $k$-body interactions: the Young diagrams, associated with the initial and the final states of a transition, can differ by relocation of no more than $k$ boxes between their rows; and (b) correlation rules: eigenstate-averaged local correlations of $k$ particles vanish if $k$ exceeds the number of columns (for bosons) or rows (for fermions) in the associated Young diagram. It also elucidates the physical meaning of the quantities conserved due to permutation symmetry --- in 1929, Dirac identified those with characters of the symmetric group --- relating them to experimentally observable correlations of several particles.
 
The results provide a way to control the formation of entangled states belonging to multidimensional non-Abelian representations of the symmetric group. These states can find applications in quantum computation and metrology.
\end{abstract}


\maketitle
Selection rules constrain possible transitions between states of quantum systems \cite{dirac,landau}. They allow the prediction of essential properties of physical systems based on their symmetries, without expensive calculations. Like every other symmetry, permutation symmetry leads to conservation laws, which were identified by Dirac\cite{dirac1929} (see also\cite{dirac}). This symmetry has been used in the Yang-Gaudin model \cite{yang1967,*sutherland1968} and has gained increasing attention due to recent progress in the control of many-body states of cold atoms \cite{fang2011,daily2012,harshman2014}. 

The Pauli exclusion principle (see the review \cite{kaplan2013} and references therein)  states that a many-body wavefunction changes its sign on permutation of two identical fermions and remains unchanged on permutation of two identical bosons.
At first glance, this fixes the permutation properties of each system and leave no room for selection rules.
However, the symmetric group $\pr{S}_N$ of permutations of $N$ symbols has also multidimensional, non-Abelian, irreducible representations (irreps), when a permutation  operator $\pr{P}$ transforms the wavefunction into a superposition of several wavefunctions in the representation (see \cite{hamermesh,kaplan,pauncz_symmetric}). 
In physical systems, such wavefunctions can appear where   
a many-body  Hamiltonian $\hat{H}=\hat{H}_{\mathrm{spat}}+\hat{H}_{\mathrm{spin}}$ is a sum of a  spin-independent $\hat{H}_{\mathrm{spat}}$ and coordinate-independent $\hat{H}_{\mathrm{spin}}$, and each of $\hat{H}_{\mathrm{spat}}$ and $\hat{H}_{\mathrm{spin}}$ is permutation-invariant. For example, $\hat{H}_{\mathrm{spat}}$ can represent particles with spin-independent interactions, and $\hat{H}_{\mathrm{spin}}$ can describe an interaction with a homogeneous magnetic field. 
The spatial and spin eigenfunctions of $\hat{H}_{\mathrm{spat}}$ and $\hat{H}_{\mathrm{spin}}$, respectively, form multidimensional irreps of the symmetric group. The total wavefunction is a sum of products of the spin and spatial functions and
satisfies the exclusion principle. 
Hamiltonians and wavefunctions of this type appear in the spin-free quantum chemistry \cite{pauncz_symmetric}. 
They can also describe spinor quantum gases, which are extensively studied starting from the first experiments \cite{myatt1997,stamper1998} and the classical theoretical investigations \cite{ho1998,*ohmi1998} (see book \cite{pitaevskii}, reviews \cite{stamper2013,*guan2013}, and references therein). Such gases, containing atoms in several states (hyperfine or magnetic), can demonstrate a variety of non-trivial symmetries (see \cite{wu2003,*wu2006} and references therein). 
A general Hamiltonian of a spinor gas \cite{ho1998} contains spin-dependent interactions. However, if atoms have closed electron shells and nuclear spins (e.g., ${}^{87}$Sr \cite{boyd2006,*desalvo2010,*tey2010} and ${}^{173}$Yb \cite{fukuhara2007},  used in experiments), the interactions will be spin-independent with a good accuracy due to weak interaction of nuclear magnetic moments.
Spin-independent interactions between the atoms can also be provided by magnetic, optical, or microwave Feshbach resonances (see \cite{stamper2013,*guan2013} and references therein). In these cases, the Hamiltonian can be separated to spin-independent and coordinate-independent parts. Instead of coordinates and spins, other two kinds degrees of freedom can be considered, e.g., electronic and spin ones \cite{gorshkov2010}.

\begin{figure}
 \includegraphics[width=3.4in]{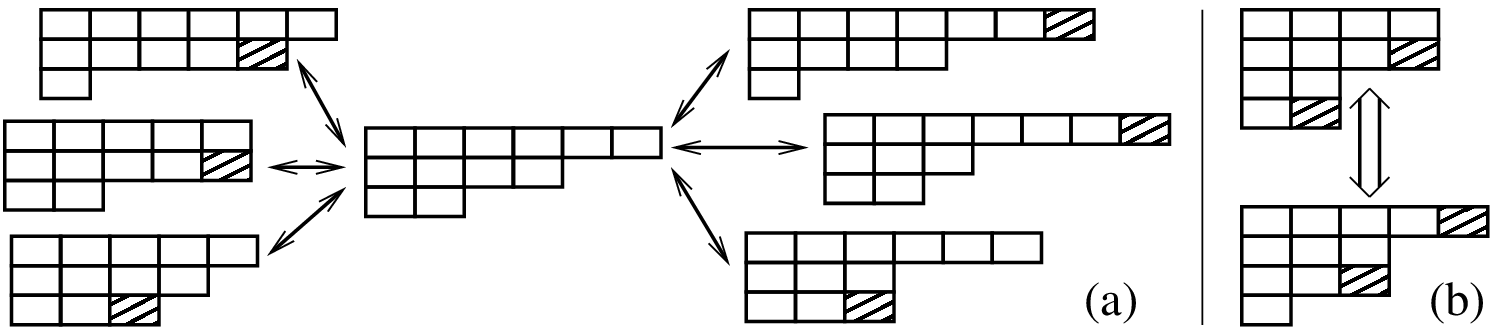}
 \caption{Selection rules \protect\refneqs{SelRule}.
 (a) Six states, which can be coupled to a given state (associated with the central Young diagram) by a one-body interaction $\hat{W}_1$ [\protect\refeq{Wk}] for $s=1$. 
 (b) An example of the coupling by a two-body interaction $\hat{W}_2$ [\protect\refeq{Wk}]. The dashed boxes are relocated. \label{Fig_sel}}
\end{figure}

If $\hat{H}_{\mathrm{spin}}$ is independent of the spin components, the gas becomes to be $SU(M)$-symmetric \cite{honerkamp2004,gorshkov2010,cazalilla2009}, where $M=2s+1$ is the multiplicity and $s$ is the spin of the atom. This symmetry has been recently observed in experiments \cite{zhang2014,*scazza2014}. States of $SU(M)$-symmetric systems are classified according to the Young diagrams $\lambda=[\lambda_1,\ldots,\lambda_M]$ --- sets of $M$ non-negative non-increasing integers $\lambda_m$ that sum to $N$ [they are pictured as $M$ rows of $\lambda_m$ boxes, see e.g. Fig.\ \ref{Fig_sel}]. 
Transformations in the spin space couple functions within irrep of $SU(M)$. Functions in different irreps, associated with the same Young diagram, are coupled by permutations of particles, forming irreps of the symmetric group. A set of all states associated with the Young diagram $\lambda$ will be referred to here as a $\lambda$-multiplet. In generic, non-$SU(M)$ invariant systems with coordinate-independent $\hat{H}_{\mathrm{spin}}$, only the permutation symmetry survives.
If $s=1/2$, the Young diagram is unambiguously determined by the total spin of the many-body system $S$ as $\lambda_1-\lambda_2=2S$. If $s>1/2$, the irreps of both groups contain contributions with different total spins \cite{landau,kaplan}. 

Every permutation $\pr{P}$ commutes with the Hamiltonian and, therefore, is an integral of motion \cite{dirac,dirac1929}. However, permutations do not commute with each other. The commuting integrals of motion \cite{dirac,dirac1929} are the character operators $\hat{\chi}(C_N)=\sum_{\pr{P}\in C_N}\pr{P}/g(C_N)$. Here
the sum is over all $g(C_N)$ permutations $\pr{P}$ of $N$ particles in a conjugate class $C_N$ (two permutations $\pr{P}$ and $\pr{P}'$ are conjugate if there exist a permutation $\pr{Q}$ such that $\pr{P}'=\pr{Q}^{-1} \pr{P}\pr{Q}$, see \cite{hamermesh,kaplan,pauncz_symmetric}). The operator $\hat{\chi}$ for transpositions (permutations of two particles) was also used \cite{fang2011} for the classification of states of a Bose-Fermi mixture.

\paragraph{Wavefunctions}
The spin $\Xi^{[\lambda]}_{t l}$ and spatial $\Phi^{[\lambda]}_{t n}$ eigenfunctions form irreps of the symmetric group, associated with the Young diagram $\lambda$, and are transformed by a permutation $\pr{P}$ as \cite{hamermesh,kaplan,pauncz_symmetric}
$
\pr{P} \Xi^{[\lambda]}_{t l}=\sum_{t'} D_{t' t}^{[\lambda]}(\pr{P})\Xi^{[\lambda]}_{t' l}
$,
$
\pr{P} \Phi^{[\lambda]}_{t n}=\mathrm{sig}(\pr{P})\sum_{t'} D_{t' t}^{[\lambda]}(\pr{P})\Phi^{[\lambda]}_{t' n}
$,
where the standard Young tableaux $t$ and $t'$ of the shape $\lambda$ label the functions within irreps, and $D_{t' t}^{[\lambda]}(\pr{P})$ are the Young orthogonal matrices (see \cite{kaplan,pauncz_symmetric}). The factor $\mathrm{sig}(\pr{P})$ is the permutation parity for fermions and $\mathrm{sig}(\pr{P})\equiv 1$ for bosons. For fermions 
$D_{t' t}^{[\tilde{\lambda}]}(\pr{P})\equiv\mathrm{sig}(\pr{P})D_{t' t}^{[\lambda]}(\pr{P})$
are matrices of the conjugate representation with $\tilde{\lambda}$ obtained from $\lambda$ by changing rows and columns. The functions belonging to the same irrep can be considered as components of a vector (or pseudovector) of the same dimension $f_\lambda$ as the representation. Each permutation corresponds then to a rotation [represented by the matrix $D_{t' t}^{[\lambda]}(\pr{P})$] of the vectors. The total wavefunction 
\begin{equation}
\Psi^{[\lambda]}_{n l}=f_\lambda^{-1/2}\sum_{t}
\Phi^{[\lambda]}_{t n}\Xi^{[\lambda]}_{t l}   ,
\label{Psilamnl}
\end{equation}
being a scalar product of the vectors of the spin and spatial wavefunctions, is then scalar (or pseudoscalar) and is transformed as
$\pr{P}\Psi^{[\lambda]}_{n l}=\mathrm{sig}(\pr{P})\Psi^{[\lambda]}_{n l}$, in the agreement to the exclusion principle. 
Different irreps, associated with the same Young diagram, are labeled by $n$ and $l$ for the spatial and spin functions, respectively.

Each particle ($j$) can occupy one of the spin states $|m(j)\rangle$, $1\leq m\leq M=2s+1$. The quantum number $m$ can also denote internal states of composite particles, e.g., hyperfine states of atoms. In the last case, the even (odd) number $M$ of internal states corresponds to the integer (half-integer) spin, with no relation to the permutation symmetry of the total wavefunction. The many-body spin eigenfunctions are expressed as sums of configurations \refsup{SI_SpinWF},
\begin{equation}
\Xi^{[\lambda]}_{t l}=\sum_{\{N\},r} B^{[\lambda]}_{l \{N\} r}
\sum_{\pr{P}}D_{t r}^{[\lambda]}(\pr{P}) 
\prod_{j=1}^{N}|m_j(\pr{P}j)\rangle . 
\label{SpinFun}
\end{equation} 
Here the configurations correspond to the different occupations $N_m$ of the states $|m\rangle$, such that $m_j=m$ for 
$\sum_{i=1}^{m-1}N_i< j\leq\sum_{i=1}^{m}N_i$.
The spatial wavefunction can be represented in a similar form \refsup{SI_SpatWF}, like the configuration-interaction method in quantum chemistry (see \cite{pauncz_symmetric}). The total wavefunction \refneq{Psilamnl} cannot be represented as a product of the states of individual particles. It is therefore a wavefunction of a many-body entangled state.

The spin state occupations $N_m$ in the wavefunction \refneq{SpinFun} are restricted by the associated Young diagram.
For $s=1/2$-particles, the total spin $S=\lambda_1-N/2$. Its projection $S_z$ is related to the occupations $N_{\uparrow\downarrow}$ of the spin up/down states as $S_z=N_{\uparrow}-N/2=N/2-N_{\downarrow}$, leading to $N_{\uparrow\downarrow}\leq \lambda_1$, since $-S\leq S_z\leq S$. Similar restrictions are obtained in the general case of $s>1/2$ \cite{supplement}. They are:
the spin-state occupations $N_i$ cannot exceed the first row length, $N_i\leq \lambda_1$ (obtained in \cite{kaplan}); if occupations of $m$ states ($1\ldots m$ for definiteness) are equal to lengths of the first $m$ rows, $N_i=\lambda_i (1\leq i \leq m)$, occupations of other states cannot exceed the next row length, $N_i\leq\lambda_{m+1} (m+1\leq i\leq N$). This demonstrates the physical meaning of the Young diagrams. These restrictions are valid for spatial functions as well; for fermions the spatial state occupations are restricted by row lengths of the conjugate Young diagram $\tilde{\lambda}$, which are equal to the column lengths of $\lambda$.

\begin{figure}
 \includegraphics[width=3.4in]{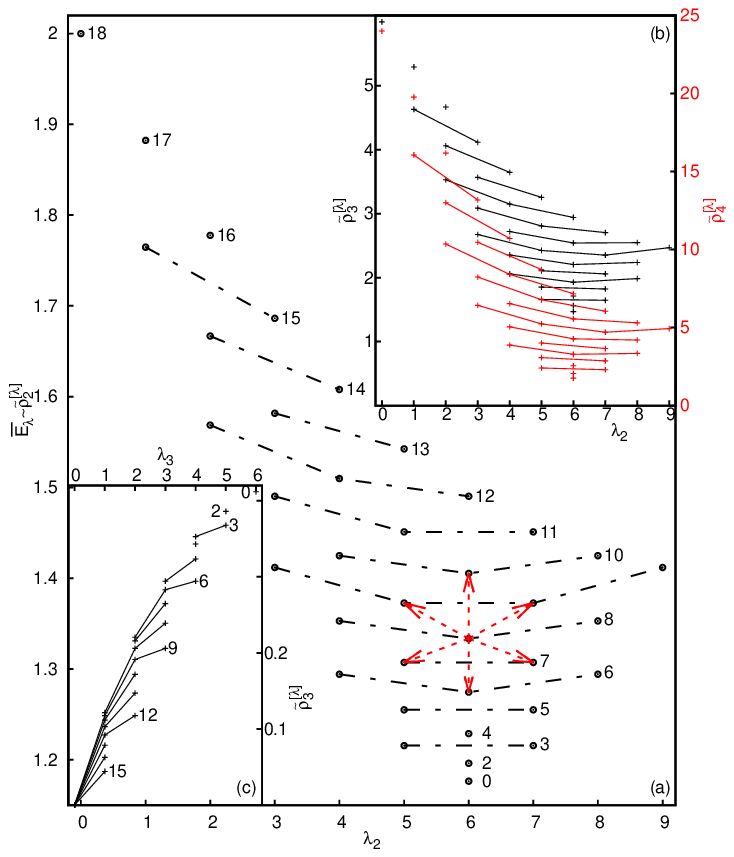}
 \caption{
 (a) Allowed transitions between 
 $\lambda=[(N-\lambda_2+S)/2,\lambda_2,(N-\lambda_2-S)/2]$-multiplets (red arrows) due to a one-body interaction $W_1$ \protect\refneqs{Wk} for $N=18$ bosons with the spin $s=1$. The average energies of the multiplets [circles, see \protect\refeq{Eav0}] for bosons are proportional to the average local two-body correlations \protect\refneq{rhotilklam}. The dashed lines connect the points with the same maximal spin $S$.
 (b) The $3$- and $4$-body  average local correlations for the same multiplets (black and red, respectively).
 (c) The $3$-body  average local correlations for the $N=18$ 
 fermions
 with the spin $s=1$ as functions of $\lambda_3$ given the maximal spin $S$ (denoted by numbers).   \label{Fig_sel_en_cor}}
 \end{figure}

\paragraph{Selection rules}
If an interaction depends on spins or coordinates only, it can couple only the states \refneqs{Psilamnl} associated with the same Young diagram, due to orthogonality of the spatial or spin functions, respectively. A nonseparable spin- and coordinate-dependent interaction of $k$ particles $j_1,\ldots,j_k$,
\begin{eqnarray}
 \hat{W}_k(\{j\})=\sum_{\{m\},\{m'\}}\langle\{m'\}|\hat{W}(\mathbf{r}_{j_1},\ldots,\mathbf{r}_{j_k})|\{m\}\rangle
 \nonumber
\\
 \times
\prod_{i=1}^k|m'_i(j_i)\rangle\langle m_i(j_i)| ,
 \label{Wk}
\end{eqnarray} 
can couple only the states if their Young diagrams, $\lambda$ and $\lambda'$, differ by relocation of no more than $k$ boxes between their rows \refsup{SI_sel}. 
These selection rules (see Fig.\ \ref{Fig_sel}) can be expressed as
\begin{equation}
 \sum_{m=1}^M |\lambda_m-\lambda'_m|\leq 2k  ,
 \label{SelRule}
\end{equation} 
while both diagrams have to satisfy the standard relations, $\lambda_{m+1}\leq\lambda_m$, $\lambda_m\geq 0$, and $\sum_{m=1}^M\lambda_m=N$. For $s=1/2$ and $k=1$, we have 
$|\lambda'_1-\lambda_1|\leq 2$, or $|S'-S|\leq 1$. It agrees to the conventional selection rule for dipole transitions. Although many-body states of higher-spin particles generally do not have a defined total spin \cite{landau,kaplan}, a maximal spin
\begin{equation}
S=(s+1)N-\sum_{m=1}^M m\lambda_m 
\label{MaxSpin}
\end{equation}  
can be introduced \refsup{SI_MaxSpin}. However, in this case the selection rule restricts $2s$ parameters and cannot be expressed in therms of $S$ alone. For example, for $s=1$ and $k=1$ there are only 6 allowed transitions [see Figs.\ \ref{Fig_sel}(a) and \ref{Fig_sel_en_cor}(a)],  although the allowed number of $\lambda$-multiplets with given $S$ is of order of $N$.

\paragraph{Correlation rules}
The probabilities of finding the given distances $\mathbf{R}_i$ between $k$ particles or the given differences $\mathbf{q}_i$ between their momenta, the $k$-body spatial $\bar{\rho}_k(\{\mathbf{R}\})$ or  momentum  $\bar{g}_k(\{\mathbf{q}\})$ correlations, respectively, are the expectation values of the operators 
\begin{eqnarray}
\hat{\rho}_k(\{\mathbf{R}\})=\prod_{i=2}^k
\delta(\mathbf{r}_1-\mathbf{r}_i-\mathbf{R}_{i-1}) 
\label{hatrhok}
\\
\hat{g}_k(\{\mathbf{q}\})=\prod_{i=2}^k
\delta(\mathbf{p}_1-\mathbf{p}_i-\mathbf{q}_{i-1})  .
\label{hatgk}
\end{eqnarray} 
Here $\mathbf{r}_i$ and  $\mathbf{p}_i$ are, respectively, $D$-dimensional coordinates and momenta (in physical applications, $D$ can be either $1$, $2$, or $3$), and for $D>1$ the $\delta$-functions in \refneq{hatrhok} are properly renormalized.
The local correlations, probabilities of finding $k$ particles in the same point (or with the same momenta) are determined by $ \hat{\rho}_k(\{0\})$ (or $ \hat{g}_k(\{0\})$). Their eigenstate expectation values, 
$\langle \Psi^{[\lambda]}_{n l}|\hat{\rho}_k(\{0\})|\Psi^{[\lambda]}_{n l}\rangle$ and $\langle \Psi^{[\lambda]}_{n l}|\hat{g}_k(\{0\})|\Psi^{[\lambda]}_{n l}\rangle$,
vanish if the correlation order $k$ exceeds the first row length in the Young diagram for the spatial wavefunction --- $\lambda_1$ for bosons or $\tilde{\lambda}_1$ for fermions (which is equal to the number of rows in the Young diagram $\lambda$ for the spin wavefunction)\cite{supplement}. 
For fermions, these restrictions are stricter than the ones provided by the Pauli principle, which states that $k$ cannot exceed the number of different spin states (this number can be greater than the number of rows).

\paragraph{Correlations and characters}
The $\lambda$-multiplet-average of a $k$-body spin-independent operator $\hat{F}_k$ is expressed as \refsup{SI_MultAv},
 \begin{eqnarray}
&& \bar{F}^{[\lambda]}_k\equiv\frac{1}
 {\tilde{\mathcal{N}}_{\lambda}}
 \sum_n \langle \Psi^{[\lambda]}_{n l}|\hat{F}_k|\Psi^{[\lambda]}_{n l}\rangle
  \nonumber
  \\
&& =\frac{f_{\lambda}}{\tilde{\mathcal{N}}_{\lambda}}
 \sum_{C_N}\mathrm{sig}(C_N)g(C_N)\tilde{\chi}_{\lambda}(C_N)
 \langle F_k\rangle_{C_N}
 \label{Fkexp}
 \end{eqnarray}
where $\tilde{\mathcal{N}}_{\lambda}$
is the total number of the spatial wavefunctions, associated with the Young diagram $\lambda$. The multiplet-dependence is given by the normalized characters $\tilde{\chi}_{\lambda}(C_N)$. The factors 
$\langle F_k\rangle_{C_N}$ \refsup{SI_MultAv} are independent of the multiplet.
If $\hat{F}_k$ is the coordinate-dependent Hamiltonian $\hat{H}_{\mathrm{spat}}$ and each spatial orbital is occupied only by one particle, \refeq{Fkexp} is reduced to the average multiplet energy, obtained in \cite{heitler1927}. 

The local spatial (or momentum) correlations are determined by $\langle \rho_k(\{0\})\rangle_{C_k}$ (or $\langle g_k(\{0\})\rangle_{C_k}$), which become independent  of the conjugate class $C_k$ \refsup{SI_corr} if each spatial orbital is occupied only by one particle. In this case 
$\tilde{\mathcal{N}}_{\lambda}=f_{\lambda}$. Then the multiplet dependence of the average local correlations, $\bar{\rho}^{[\lambda]}_k(\{0\})=\tilde{\rho}^{[\lambda]}_k\langle \rho_k(\{0\})\rangle$ and $\bar{g}^{[\lambda]}_k(\{0\})=\tilde{\rho}^{[\lambda]}_k\langle g_k(\{0\})\rangle$,  is given by the universal factor
\begin{equation}
 \tilde{\rho}^{[\lambda]}_k=
 \sum_{C_k}\mathrm{sig}(C_k)g(C_k)\tilde{\chi}_\lambda (C_k)  .
 \label{rhotilklam}
\end{equation} 
Thus, the integrals of motion $\tilde{\chi}_\lambda (C_k)$, corresponding to the permutation symmetry, are related to quantities $\bar{\rho}^{[\lambda]}_k(\{0\})$ and $\bar{g}^{[\lambda]}_k(\{0\})$, which can be measured in experiments.

In a system with zero-range two-body interactions, 
$V(\mathbf{r}'-\mathbf{r})=V N(N-1)\hat{\rho}_2(\{0\})/2$,
the average energy of the $\lambda$-multiplet, counted from the multiplet-independent energy of non-interacting particles, is
\begin{equation}
 \bar{E}_{\lambda}=V\frac{N(N-1)}{2}[1\pm \tilde{\chi}_\lambda (\{2\})]\langle \rho_2(\{0\})\rangle  .
 \label{Eav0}
\end{equation} 
Here, the sign $+/-$ is taken for bosons/fermions and $\{2\}$ is the conjugate class of the transpositions \refsup{SI_MultAv}. The energy attains its maximum for bosons and minimum for fermions at $\lambda=[N]$, when
the normalized character for transpositions attains its maximum $\tilde{\chi}_\lambda (\{2\})=1$  \cite{supplement}. In this state (belonging to a one-dimensional irrep) the total spin is defined and has the maximal allowed value $N s$. The minimal average energy for bosons and the maximal one for fermions correspond to 
$\lambda=[(\lambda_M+1)^k,\lambda_M^{M-k}]$, where  $\tilde{\chi}_\lambda (\{2\})$ attains its minimum  $[(N+k)\lambda_M+(M-k+1)k-M N]/[N(N-1)]$ \cite{supplement}. (Here $\lambda_M$ and $k$ are, respectively, the quotient and remainder of the division of $N$ by $M$.) This $\lambda$-multiplet  corresponds to the minimum $S=(M-k)k/2$ of the maximal spin. If $N$ is a multiple of $M$, it has the defined total spin $S=0$.

These general properties are confirmed for particular values of $s$ using the explicit expressions \cite{supplement} obtained with the characters \cite{lassalle2008}.
For $s=1/2$, the energy 
$\bar{E}_{\lambda}=V[N(N-1)/2\pm (N^2/4-N+S^2+S)]\langle \rho_2(\{0\})\rangle$
is a monotonic function of the total spin $S$. For $s=1$, $\bar{E}_{\lambda}=V[N(N-1)/2 \pm(N^2-6N+S^2+4S+3\lambda_2^2-2N\lambda_2)/4]\langle \rho_2(\{0\})\rangle$
is a sum of quadratic functions of the maximal spin $S$ and $\lambda_2$.
Its  dependence of $S$ may be non-monotonic [see Fig.\ \ref{Fig_sel_en_cor}(a)]. The multiplet-dependencies of the $3$- and $4$-body correlations ($\tilde{\rho}^{[\lambda]}_3$ and $\tilde{\rho}^{[\lambda]}_4$) are shown in Fig.\ \ref{Fig_sel_en_cor}(b,c). For fermions, $3$-body correlations vanish for two-row Young diagrams [$\lambda_3=0$, see Fig.\ \ref{Fig_sel_en_cor}(c)], in agreement with the correlation rules. The averages are independent of the particle spin, which only restricts the number of the Young diagram rows. 

 \begin{figure}
 \includegraphics[width=3.4in]{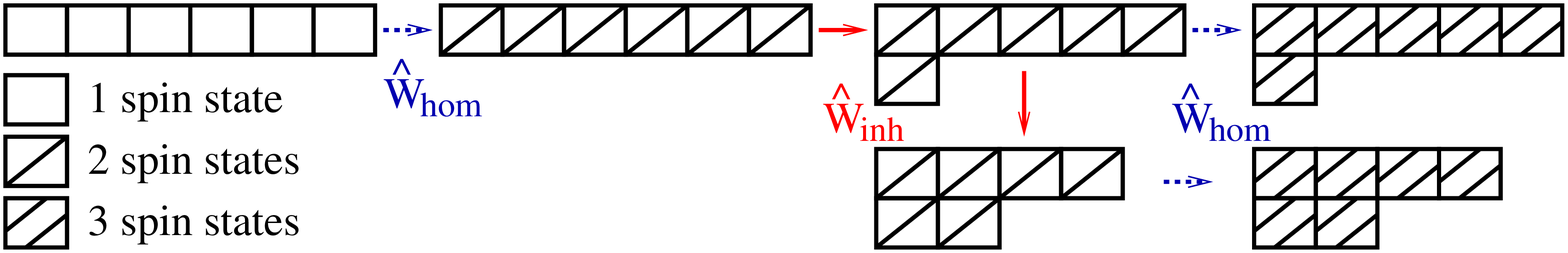}
 \caption{A scheme of population of $\lambda$-multiplets using spatially-homogeneous spin-changing pulses $\hat{W}_{\text{hom}}(t)$ (blue dashed arrows) and spatially-inhomogeneous spin-conserving pulse $\hat{W}_{\text{inh}}(\mathbf{r},t)$ (red solid arrows). Shapes of the Young diagram boxes denote numbers of occupied atomic spin states. \label{Fig_trans}}
 \end{figure}

\paragraph{Possible realization}
The states, associated with various Young diagrams, may be selectively populated using two types of pulses (see Fig.\ \ref{Fig_trans}). A spatially-homogeneous spin-changing pulse
$\hat{W}_{\text{hom}}(t)=\sum_{m\neq m'}W_{m m'}(t)|m\rangle\langle m'|$
changes the spin states of atoms but, being coordinate-independent, does not change the Young diagram associated with the many-body state. $\pi/2$ pulses of this type are generally used in experiments with spinor cold gases \cite{matthews1998,sagi2010}.
A spatially-inhomogeneous spin-conserving pulse
$\hat{W}_{\text{inh}}(\mathbf{r},t)=\sum_{m}W_{m}(\mathbf{r},t)|m\rangle\langle m|$
is a one-body interaction of the form \refneq{Wk}. It can relocate one box in the Young diagram, according to the selection rules, but does not change the spin states of the atoms. If all atoms are initially formed in the same spin states, the many-body spin wavefunction is associated with the one-row Young diagram $[N]$. A pulse of the type $\hat{W}_{\text{hom}}(t)$ can transfer each atom to a superposition of two spin states. For fermions, all local correlations vanish in this state, since its spatial wavefunction is associated with the one-column Young diagram. Then a pulse of the type 
$\hat{W}_{\text{inh}}(\mathbf{r},t)$ can lead to the spin wavefunction associated with a two-row Young diagram $[N-1,1]$, depleting the $[N]$ state. The depletion could be detected in a Ramsey experiment (like \cite{sagi2010,sagi2010a}) by applying the second  pulse $\hat{W}_{\text{hom}}(t)$. Further pulses of the type $\hat{W}_{\text{inh}}(\mathbf{r},t)$ can provide only one- and two-row Young diagrams, since only two atomic spin states are occupied. For fermions, only two-body local correlations do not vanish in these many-body states. A population of the third atomic spin state by $\hat{W}_{\text{hom}}(t)$ does not change the vanishing correlations, but allows to provide three-row Young diagrams using $\hat{W}_{\text{inh}}(\mathbf{r},t)$. States, associated with arbitrary Young diagrams can be populated in this way, and, for fermions, the number of rows can be tested by non-vanishing correlations. 
More comprehensive information on the populated states can be provided by the correlation values, since they are related to the characters [see \refneq{rhotilklam}] and the Young diagram is unambiguously related to the values of all characters \cite{dirac,dirac1929}. Then the correlations can allow to analyze coherent or statistical mixtures of various $\lambda$-multiplets.

The selection and correlation rules are applicable to any system with two kinds of separable degrees of freedom, e.g. to a spinor gas with spin-independent interactions in arbitrary trap potentials. The simple relation \refneq{rhotilklam} between correlations and characters is obtained for the single occupations of spatial orbitals. This can be realized, for example, with cold atoms in a $D$-dimensional optical lattice \cite{bloch2008,*yukalov2009,*svistunov} in the unit-filling Mott (or fermionic band)-insulator regime, when each lattice site is occupied by one atom.
In this regime, the spatial correlations do not demonstrate a substantial dependence on the spin state \refsup{SI_latt} (indeed, in any state, each site is occupied by one atom). The momentum correlations oscillate as functions of each component of $\mathbf{q}_j$ with the maximal values (at $\mathbf{q}_j=0$)  \refsup{SI_latt} 
\begin{equation}
 \bar{g}_k(\{0\})=\tilde{\rho}^{[\lambda]}_k f_k(\{0\}) .
 \label{bargkzero}
\end{equation}
Here the probability of finding differences $\mathbf{q}_{j}$ between momenta of $k$ non-interacting particles 
$
 f_k(\{\mathbf{q}\})=\int d^D p |\tilde{w}(\mathbf{p})|^2
 \prod_{i=2}^k |\tilde{w}(\mathbf{p}+\mathbf{q}_{i-1})|^2
$ 
is the convolution of the momentum distributions 
$|\tilde{w}(\mathbf{p})|^2$. 
The correlations and momentum distributions in \refneq{bargkzero} can be measured in experiments, and the factor $\tilde{\rho}^{[\lambda]}_k$ is a linear combination of the characters \refneq{rhotilklam}.

\paragraph{Conclusions} 
Rather abstract mathematical constructs --- Young diagrams and characters of the symmetric group --- have a physical meaning. Young diagrams classify many-body states of systems with separable spin and spatial degrees of freedom. For such state, a maximal spin \refneq{MaxSpin}, occupations of one-body states, and non-vanishing correlations are determined by row lengths and number of rows in the associated Young diagram. A transition due to a nonseparable $k$-body interaction cannot move more than $k$ boxes between the Young diagram rows [see selection rules \refneq{SelRule}]. The characters ---  integrals of motion, corresponding to permutation symmetry --- are related to correlations of several particles in the coordinate or momentum space [see  \refneq{Fkexp} and \refneq{rhotilklam}], which can be measured in experiments. This demonstrates that the characters have a physical meaning, similarly to the integrals of motion corresponding to many other symmetries.

This research was supported in part by the National Science Foundation under Grant No. NSF PHY11-25915.
The author gratefully acknowledge useful conversations with A. Ben-Reuven, N. Davidson, I. G. Kaplan, M. Olshanii, R. Pugatch, A. Simoni, and B. Svistunov.

%

\clearpage

\renewcommand{\theequation}{S-\arabic{equation}}
\setcounter{equation}{0}
\begin{widetext}
\begin{center}
{\Large \bf Supplemental material for: Permutation symmetry in spinor quantum gases: selection rules, conservation laws, and correlations}

Vladimir A. Yurovsky

\end{center}
\end{widetext}

\tableofcontents

\bigskip 

Numbers of equations in the Supplemental material are started from S. References to equations in the Letter do not contain S.

\section{Spin wavefunctions}\label{SI_SpinWF}
Suppose that $N$ particles, labeled by $j$, occupy $M$ orthonormal one-body states $|m(j)\rangle$, $1\leq m\leq M$. The basic functions of irreducible representations (irreps) of the symmetric group $\pr{S}_N$ can be expressed as (see \cite{kaplan,pauncz_symmetric}),
\begin{equation}
 |\{N\},\lambda,t,r\rangle=\left( \frac{f_{\lambda}}{N!}\right)^{1/2}
 \sum_{\pr{P}}D_{t r}^{[\lambda]}(\pr{P})\Xi_{\{N\}\pr{P}} ,
 \label{Nlamtr}
\end{equation} 
where $D_{t r}^{[\lambda]}(\pr{P})$ are the Young orthogonal matrices, $\lambda$ is a Young diagram, associated with the irrep, $t$, $r$ are the standard Young tableaux of the shape $\lambda$, and $\pr{P}$ are permutations of $N$ symbols. A permutation $\pr{P}$ transforms the functions \refneqs{Nlamtr} as
\[
 \pr{P}|\{N\},\lambda,t,r\rangle =\sum_{t'}D_{t' t}^{[\lambda]}(\pr{P})
 |\{N\},\lambda,t',r\rangle.
\] 
Thus $t$ labels the basic functions of the representation and $r$ labels different representations, associated with the same $\lambda$. The irrep dimension $f_{\lambda}$ is equal to the number of standard Young tableaux of the shape $\lambda$ \cite{kaplan},
\[
 f_{\lambda}=\frac{N!\prod_{m<m'}(\lambda_m-m-\lambda_{m'}+m')}
 {\prod_{m=1}^M (\lambda_m+M-m)!} .
\] 
The permutted states
\begin{equation}
 \Xi_{\{N\}\pr{P}}=\prod_{j=1}^N |m_j(\pr{P}j)\rangle
 \label{XiNP}
\end{equation} 
are determined by the set $\{N\}$ of occupations $N_m$ of the states 
$|m\rangle$. 
The one-body states are arranged in non-decreasing order of $m$, such that $m_j=m$ for $\sum_{i=1}^{m-1}N_i< j\leq\sum_{i=1}^{m}N_i$ and $\sum_{m=1}^{M}N_m=N$. The number of possible sets $\{N\}$, the number $\mathcal{N}(N,M)$ of distributions of $N$ identical particles to $M$ distinct states, is calculated in combinatorics \cite{bogart} as
\[
 \mathcal{N}(N,M)=\frac{(N+M-1)!}{N!(M-1)!} .
\] 

If all states have single occupation ($N_m=1$), all $m_j$ are different, all the functions \refneqs{Nlamtr} are orthogonal,
\[
 \langle \{N\},\lambda',t',r'|\{N\},\lambda,t,r\rangle =
 \delta_{\lambda \lambda'} \delta_{t t'} \delta_{r r'},
\] 
and $r$ can be any of the $f_{\lambda}$ standard Young tableaux. However, if multiple occupations of the states $|m\rangle$ are allowed, the permutted states \refneqs{XiNP} become non-orthogonal
\[
 \langle \Xi_{\{N'\}\pr{P'}}|\Xi_{\{N\}\pr{P}}\rangle=\delta_{\{N\} \{N'\}}
 \sum_{\pr{P}^{\{N\}}}\delta_{\pr{P'},\pr{P}\pr{P}^{\{N\}}}  ,
\] 
where 
\begin{equation}
\pr{P}^{\{N\}}=\prod_{m=1}^M\pr{P}^{(m)}
\label{PN}
\end{equation} 
and $\pr{P}^{(m)}$ can be either the identity permutation or any permutation of $\sum_{i=1}^{m-1}N_i< j\leq\sum_{i=1}^{m}N_i$, leaving all other $j$ unchanged.

This leads to non-orthogonality of the basic functions \refneqs{Nlamtr} in different irreps,
 \begin{eqnarray}
 \langle \{N'\},\lambda',t',r'&|&\{N\},\lambda,t,r\rangle =
 \delta_{\{N\} \{N'\}}\frac{\sqrt{f_{\lambda'}f_{\lambda}}}{N!}
 \nonumber
 \\
 &&\times
\sum_{\pr{P},\pr{P}^{\{N\}}}D_{t' r'}^{[\lambda']}(\pr{P})
 D_{t r}^{[\lambda]}(\pr{P}\pr{P}^{\{N\}})
\nonumber
\\
&& = \delta_{\{N\} \{N'\}}\delta_{\lambda \lambda'} \delta_{t t'} 
 \tilde{D}_{r' r}^{[\lambda]}(\{N\}),
 \phantom{qqqq}
 \label{Nlamtrover}
\end{eqnarray} 
where
\begin{equation}
\tilde{D}_{r' r}^{[\lambda]}(\{N\})=
\sum_{\pr{P}^{\{N\}}}D_{r' r}^{[\lambda]}(\pr{P}^{\{N\}}) .
\label{tildeD}
\end{equation} 
The above derivation uses the following properties of the Young orthogonal matrices (see \cite{kaplan,pauncz_symmetric}),
\begin{eqnarray}
 D_{r t}^{[\lambda]}(\pr{P}\pr{Q})=
 \sum_{t'} D_{r t'}^{[\lambda]}(\pr{P}) D_{t' t}^{[\lambda]}(\pr{Q})
 \label{ProdYoung}
 \\
 \sum_{\pr{P}} D_{t' r'}^{[\lambda']}(\pr{P})D_{t r}^{[\lambda]}(\pr{P})
= \frac{N!}{f_{\lambda}}\delta_{\lambda \lambda'} \delta_{t t'} \delta_{r r'} 
\label{OrthYoung}
\end{eqnarray}
The matrix $\tilde{D}_{r' r}^{[\lambda]}$ is symmetric, since 
$D_{r r'}^{[\lambda]}(\pr{P})=D_{r' r}^{[\lambda]}(\pr{P}^{-1})$ for orthogonal matrices and $(\pr{P}^{\{N\}})^{-1}$ belongs to the subgroup $\{\pr{P}^{\{N\}}\}$. Therefore it has $f_{\lambda}$ orthogonal normalized eigenvectors
$d_{\nu r}^{[\lambda]}(\{N\})$, corresponding to eigenvalues $\delta_{\nu}^{[\lambda]}(\{N\})$, and can be represented as
\begin{equation}
 \tilde{D}_{r' r}^{[\lambda]}=\sum_{\nu}d_{\nu r'}^{[\lambda]}
 \delta_{\nu}^{[\lambda]}d_{\nu r}^{[\lambda]} .
 \label{Devecval}
\end{equation} 
A similar representation exists for the square of this matrix,
\begin{equation}
 (\tilde{D}^{[\lambda]})_{r' r}^2=\sum_{\nu}d_{\nu r'}^{[\lambda]}
 (\delta_{\nu}^{[\lambda]})^2 d_{\nu r}^{[\lambda]}  .
 \label{D2evecval}
\end{equation}
At the same time, using \refeq{ProdYoung} we get,
\begin{eqnarray*}
 (\tilde{D}^{[\lambda]})_{r' r}^2=\sum_{t,\pr{P}^{\{N\}},\tilde{\pr{P}}^{\{N\}}}D_{r' t}^{[\lambda]}(\pr{P}^{\{N\}})
 D_{t r}^{[\lambda]}(\tilde{\pr{P}}^{\{N\}})
 \nonumber
 \\
 =\{N\}!\tilde{D}_{r' r}^{[\lambda]}  ,
\end{eqnarray*} 
since the product $\pr{P}^{\{N\}}\tilde{\pr{P}}^{\{N\}}$ belongs to the subgroup $\{\pr{P}^{\{N\}}\}$. Here 
$\{N\}!\equiv\prod_{m=1}^N N_m!$. Then, Eqs.\ \refneq{Devecval} and \refneq{D2evecval} lead to the equality 
$(\delta_{\nu}^{[\lambda]})^2=\{N\}!\delta_{\nu}^{[\lambda]}$.
This means that $\delta_{\nu}^{[\lambda]}$ can have the value of either $0$ or $\{N\}!$. Finally, \refeq{Nlamtrover} allows to prove that the functions
 \begin{equation}
  |\{N\},\lambda,t,\nu\rangle = (\{N\}!)^{-1/2} 
  \sum_r d_{\nu r}^{[\lambda]}|\{N\},\lambda,t,r\rangle ,
  \label{Nlamtnu}
 \end{equation} 
with $\nu$ corresponding to $\delta_{\nu}^{[\lambda]}>0$, are normalized and orthogonal for different $\nu$. 
 
Thus the number of irreps $\tilde{f}_{\lambda}(\{N\})$, associated with the same $\lambda$, is equal to the number of non-zero eigenvalues $\delta_{\nu}^{[\lambda]}$. Functions in these irreps form a complete basic, since Eqs.\ \refneq{Nlamtr}, \refneq{XiNP}, \refneq{Nlamtrover}, \refneq{ProdYoung}, \refneq{OrthYoung}, and \refneq{Devecval} lead to
\begin{eqnarray}
  \sum_{\lambda,t}\sum_{\nu}{}'
  \langle \Xi_{\{N\}\pr{P}'}|\{N\},\lambda,t,\nu\rangle 
  \langle \{N\},\lambda,t,\nu|\Xi_{\{N\}\pr{P}}\rangle
 \nonumber
 \\
  =\sum_{\pr{P}^{\{N\}}}\delta_{\pr{P'},\pr{P}\pr{P}^{\{N\}}}  ,
  \phantom{qqqqq}
  \label{ResIdent}
\end{eqnarray} 
where $\sum_{\nu}'$ means the summation over all $\nu$ with $\delta_{\nu}^{[\lambda]}>0$. Equation \refneq{ResIdent} is nothing but the resolution of identity, as $\Xi_{\{N\}\pr{P}\pr{P}^{\{N\}}}$ is equal to $\Xi_{\{N\}\pr{P}}$ for each $\pr{P}^{\{N\}}$. 
 
For example, if $M=2$ (as for $s=1/2$ particles) only one irrep is associated with each $\lambda$ (see \cite{yurovsky2013}, where explicit expressions are derived for the wavefunctions in this case).
 
Consider a permutation-symmetric Hamiltonian $\hat{H}_{\mathrm{spin}}=\pr{P}^{-1}\hat{H}_{\mathrm{spin}}\pr{P}$. Using Eqs.\ \refneq{Nlamtr}, \refneq{ProdYoung}, and \refneq{OrthYoung}, we get
\begin{widetext}
 \begin{eqnarray}
 \langle \{N'\},\lambda',t',r'|\hat{H}_{\mathrm{spin}}|\{N\},\lambda,t,r\rangle 
 &=&\frac{\sqrt{f_{\lambda'}f_{\lambda}}}{N!}\sum_{\pr{P},\pr{P}'}
 D_{t' r'}^{[\lambda']}(\pr{P}')D_{t r}^{[\lambda]}(\pr{P})
 \langle \Xi_{\{N'\}\pr{P}^{-1}\pr{P}'}|\pr{P}^{-1}\hat{H}_{\mathrm{spin}}\pr{P}|\Xi_{\{N\}\pr{E}}\rangle
\nonumber
 \\
&=&\frac{\sqrt{f_{\lambda'}f_{\lambda}}}{N!}\sum_{\pr{P},\pr{Q}}
 D_{t' r'}^{[\lambda']}(\pr{P}\pr{Q})D_{t r}^{[\lambda]}(\pr{P})
 \langle \Xi_{\{N'\}\pr{Q}}|\hat{H}_{\mathrm{spin}}|\Xi_{\{N\}\pr{E}}\rangle
\nonumber
 \\
 &=&\delta_{\lambda \lambda'} \delta_{t t'} \sum_{\pr{Q}}D_{r r'}^{[\lambda]}(\pr{Q})\langle \Xi_{\{N'\}\pr{Q}}|\hat{H}_{\mathrm{spin}}|\Xi_{\{N\}\pr{E}}\rangle ,
\label{Hspinmatr}
 \end{eqnarray} 
\end{widetext}
where $\pr{E}$ is the identity permutation.
This means that the coupling of the states \refneqs{Nlamtnu} is diagonal in $\lambda$ and $t$, 
$\langle \{N'\},\lambda',t',\nu'|\hat{H}_{\mathrm{spin}}|\{N\},\lambda,t,\nu\rangle \sim \delta_{\lambda \lambda'} \delta_{t t'}$, and independent of $t$ (indeed, it is a general group-theoretical property of irrep basic functions, see \cite{kaplan}). Then the eigenfunctions of $\hat{H}_{\mathrm{spin}}$ can be expanded as
\begin{equation}
  \Xi^{[\lambda]}_{t l}=\sum_{\{N\}}\sum_{\nu}{}' A^{[\lambda]}_{l \{N\} \nu}
  |\{N\},\lambda,t,\nu\rangle  ,
\label{Xilamtl_Nlamtnu}
\end{equation} 
 where the coefficients $A^{[\lambda]}_{l \{N\} \nu}$ form eigenvectors of the Hamiltonian matrix,
\begin{eqnarray}
\sum_{\{N'\}}\sum_{\nu'}{}' 
\langle \{N\},\lambda,t,\nu|\hat{H}_{\mathrm{spin}}|\{N'\},\lambda,t,\nu'\rangle
A^{[\lambda]}_{l \{N'\} \nu'}
\nonumber
\\
=E^{[\lambda]}_{l}A^{[\lambda]}_{l \{N\} \nu}  .
\phantom{qqqqq}
\label{HAeqEA}
\end{eqnarray} 
Due to hermiticity of the matrix, its eigenvectors from a complete and orthonormal basic set,
\begin{eqnarray}
 \sum_l A^{[\lambda]}_{l \{N'\} \nu'} A^{[\lambda]}_{l \{N\} \nu}=
 \delta_{\{N\}\{N'\}} \delta_{\nu \nu'}
 \label{Acomp}
 \\
 \sum_{\{N\}}\sum_{\nu}{}' A^{[\lambda]}_{l' \{N\} \nu} A^{[\lambda]}_{l \{N\} \nu}=
 \delta_{l l'} .
 \label{Aorth}
\end{eqnarray} 
Finally, using \refeq{Nlamtnu}, one obtains \reflet{\SpinFun} with
\begin{equation}
 B^{[\lambda]}_{l \{N\} r}= \left( \frac{f_\lambda}{N! \{N\}!}\right) ^{1/2} 
  \sum_{\nu}{}' d_{\nu r}^{[\lambda]}A^{[\lambda]}_{l \{N\} \nu}  .
  \label{Blamlnr}
\end{equation}

\section{Spatial wavefunctions}\label{SI_SpatWF}
Like the spin wavefunctions, the spatial ones can be expressed in terms of basic functions of the symmetric group irreps,
\begin{equation}
\tilde{\Phi}^{[\lambda]}_{\{\tilde{N}\} t r}=
\left( \frac{f_{\lambda}}{N!}\right)^{1/2}
\sum_{\pr{P}}\mathrm{sig}(\pr{P})D_{t r}^{[\lambda]}(\pr{P}) 
\prod_{j=1}^{N} \varphi_{m_j}(\mathbf{r}_{\pr{P}j}) . 
\label{tilPhilamNtr}
\end{equation} 
Here the factor $\mathrm{sig}(\pr{P})$ is the permutation parity for fermions and $\mathrm{sig}(\pr{P})\equiv 1$ for bosons. For fermions, it provides the conjugate representation with the matrices
$D_{\tilde{t} \tilde{r}}^{[\tilde{\lambda]}}(\pr{P})=\mathrm{sig}(\pr{P})D_{t r}^{[\lambda]}(\pr{P})$, where $\tilde{\lambda}$ is obtained from $\lambda$ by changing rows and columns. (In the following, the notation $\tilde{\lambda}$ will be used for bosons as well, meaning $\tilde{\lambda}=\lambda$.) The proper orthonormal one-body spatial orbitals 
$\varphi_{m}(\mathbf{r})$ ($1\leq  m\leq M_{\mathrm{spat}}$) depend on the $D$-dimensional coordinate $\mathbf{r}$ (in concrete applications, $D$ can be either $1$, $2$, or $3$). The quantum numbers $m_j$ are determined by the set of occupations $\{\tilde{N}\}$ in the same way as in the case of the spin functions. Eigenfunctions of the permutation-symmetric Hamiltonian $\hat{H}_{\mathrm{spat}}$, 
\begin{equation}
\Phi^{[\lambda]}_{t n}=\sum_{\{\tilde{N}\},r}
\left(\{\tilde{N}\}!\right) ^{-1/2}\sum_{\nu}{}' 
A^{[\lambda]}_{n \{\tilde{N}\} \nu}
d_{\nu r}^{[\tilde{\lambda}]}(\{\tilde{N}\})
\tilde{\Phi}^{[\lambda]}_{\{\tilde{N}\} t r} ,
\label{Philamtn}
\end{equation} 
are linear combinations of the basic functions \refneqs{tilPhilamNtr}, where  the coefficients $A^{[\lambda]}_{n \{\tilde{N}\} \nu}$ are solutions of the eigenproblem of the form of \refeq{HAeqEA},
\begin{eqnarray}
\sum_{\{\tilde{N}'\},r,r'}\sum_{\nu'}{}' 
\left(  \{\tilde{N}\}!\{\tilde{N}'\}!\right) ^{-1/2}
&&d_{\nu r}^{[\tilde{\lambda}]}(\{\tilde{N}\})
\nonumber
\\
\times
\langle \tilde{\Phi}^{[\lambda]}_{\{\tilde{N}\} t r}|\hat{H}_{\mathrm{spat}}|\tilde{\Phi}^{[\lambda]}_{\{\tilde{N}'\} t r'}\rangle 
&&d_{\nu' r'}^{[\tilde{\lambda}]}(\{\tilde{N}'\})
A^{[\lambda]}_{n \{\tilde{N}'\} \nu'}
\nonumber
\\
&&=E^{[\lambda]}_{n}A^{[\lambda]}_{n \{\tilde{N}\} \nu}  .
\phantom{qqqqq}
\label{HspatAeqEA}
\end{eqnarray} 
Here the Hamiltonian matrix
\begin{eqnarray}
&&\langle \tilde{\Phi}^{[\lambda]}_{\{\tilde{N}'\} t r'}|\hat{H}_{\mathrm{spat}}|\tilde{\Phi}^{[\lambda]}_{\{\tilde{N}\} t r}\rangle=
\sum_{\pr{Q}}D_{r' r}^{[\tilde{\lambda}]}(\pr{Q})
\nonumber
\\
&&\times
\int d^{DN} r \prod_{j'=1}^N 
 \varphi^*_{m'_{\pr{Q}j'}}(\mathbf{r}_{j'})\hat{H}_{\mathrm{spat}}
 \prod_{j=1}^N  \varphi_{m_{j}}(\mathbf{r}_{j})
 \label{Hspatmatr}
\end{eqnarray} 
is derived like \refeq{Hspinmatr}.

Although \refeq{Philamtn} involves only a finite number of the orbitals, the spatial wavefunction can be approximated with the required accuracy if the number of the orbitals is sufficiently large.

\section{The maximal state occupations}\label{SI_MaxOccup}
Since the permutations $\pr{P}^{\{N\}}$ \refneqs{PN} do not affect the functions $\Xi_{\{N\}\pr{P}}$ \refneqs{XiNP}, the wavefunctions \refneqs{Nlamtr} can be expressed as,
\begin{eqnarray*}
 |\{N\}&,&\lambda,t,r\rangle=\frac{1}{\{N\}!}\left( \frac{f_{\lambda}}{N!}\right)^{1/2}
 \sum_{\pr{P},\pr{P}^{\{N\}}}D_{t r}^{[\lambda]}(\pr{P})\Xi_{\{N\}\pr{P}\pr{P}^{\{N\}}}
 \nonumber
 \\
 &&=\frac{1}{\{N\}!}\left( \frac{f_{\lambda}}{N!}\right)^{1/2}
 \sum_{\pr{P},r_M}D_{t r_M}^{[\lambda]}(\pr{P})\Xi_{\{N\}\pr{P}}
 \nonumber
 \\
 &&\times
 \sum_{\pr{P}^{(M)},r_{M-1}}D_{r_M r_{M-1}}^{[\lambda]}(\pr{P}^{(M)})
 \cdots \sum_{\pr{P}^{(1)}}D_{r_1 r}^{[\lambda]}(\pr{P}^{(1)})  .
 \phantom{qq}
\end{eqnarray*} 
As $\pr{P}^{(1)}$ are elements of the subgroup $\pr{S}_{N_1}$ of permutations of $N_1$ first symbols, a reduction to subgroup (see \cite{kaplan}) 
can be used, $D_{r_1 r}^{[\lambda]}(\pr{P}^{(1)})=D_{\bar{r}_1 \bar{r}}^{[\bar{\lambda}]}(\pr{P}^{(1)})$, where
the Young tableaux $\bar{r}_1$ and $\bar{r}$, corresponding to the same Young diagram, $\bar{\lambda}$, 
are obtained by the removal of the symbols $N_1+1\ldots N$ from the tableaux $r_1$ and $r$, respectively. 
($D_{r_1 r}^{[\lambda]}(\pr{P}^{(1)})=0$ if $\bar{r}_1$ and $\bar{r}$ correspond to different Young diagrams or if the symbols $N_1+1\ldots N$ are placed differently in $r_1$ and $r$.)
Let us introduce the notation $[0]$ for the Young tableau of the proper shape in which 
the symbols are arranged by rows in the sequence of natural numbers.
Taking into account that $D_{[0] [0]}^{[N_1]}(\pr{P}^{(1)})=1$ as the Young diagram $[N_1]$, having one row
of length $N_1$, corresponds to the identity representation, we get (using \refeq{OrthYoung}),
\begin{eqnarray}
\sum_{\pr{P}^{(1)}} D_{r_1 r}^{[\lambda]}(\pr{P}^{(1)})
&=& \sum_{\pr{P}^{(1)}} D_{\bar{r}_1 \bar{r}}^{[\bar{\lambda}]}(\pr{P}^{(1)})
D_{[0] [0]}^{[N_1]}(\pr{P}^{(1)})
\nonumber
\\
&=&N_1! \delta_{\bar{\lambda} [N_1]} \delta_{\bar{r}_1 [0]} \delta_{\bar{r} [0]} . 
\label{sumP1}
\end{eqnarray}
This means that $N_1$ cannot exceed $\lambda_1$ (another proof of this statement is given in \cite{kaplan}). Besides, $r_1=r$, since, according to \refeq{sumP1} $\bar{r}_1=\bar{r}$ and the remaining parts of $r_1$ and $r$ coincide, as mentioned above.

Consider now the case of $N_1=\lambda_1$. Each permutation $\pr{P}^{(2)}$ can be represented as a product of elementary transpositions $\pr{P}_{j,j+1}$ (see \cite{kaplan,pauncz_symmetric}) of symbols $N_1< j < N_1+N_2$. All these transpositions do not affect the first row of $r$ occupied by the first $N_1$ symbols. Therefore, the Young orthogonal matrix 
$D_{r_2 r_1}^{[\lambda]}(\pr{P}^{(2)})=D_{r_2 r}^{[\lambda]}(\pr{P}^{(2)})$ will have non-zero elements only if the first row of $r_2$ is occupied by the first $N_1$ symbols. Further, as the Young orthogonal matrix for an elementary transposition depends only on the distance between the permutted symbols (see \cite{kaplan,pauncz_symmetric}), $D_{r_2 r}^{[\lambda]}(\pr{P}^{(2)})=D_{r''_2 r''}^{[\lambda'']}(\pr{P}^{(2)})$, where $r''$ and $r''_2$, obtained by removal of the first row from the tableaux $r$ and $r_2$, respectively, correspond to the same Young diagram $[\lambda'']$. The same argumentation as for $\pr{P}^{(1)}$ leads then to the conclusion that $N_2\leq \lambda_2$, the symbols $N_1+1,\ldots,N_1+N_2$ occupy the second row of $r$ in the sequence of natural numbers, and $r_2=r$. Repeating this for $D_{r_i r_{i-1}}^{[\lambda]}(\pr{P}^{(i)})$ with $3\leq i\leq m$, one gets that if $N_i=\lambda_i$ for all $1\leq i\leq m-1$ then $N_m\leq \lambda_m$, $r_m=r$, and symbols $1\leq j\leq\sum_{i=1}^{m}N_i$ occupy first $m$ rows of $r$ in the sequence of natural numbers. Therefore, there is only one irrep for $N_m=\lambda_m$ ($1\leq m\leq M$), and its label is $r=[0]$. 

\section{The maximal spin and boundaries of characters}\label{SI_MaxSpin}
Let us attribute the spin projection $s_z=s+1-m$ to the state $|m\rangle$. The functions of the irrep considered in the previous section have the maximal possible occupation  $N_1=\lambda_1$ of the state with the maximal spin projection $s$, and the occupation of the state $|m\rangle$ does not exceed occupations of the states $|m'\rangle$ with higher projections. Therefore, the functions have the maximal possible projection of the total spin. This projection 
\begin{equation}
 S=\sum_{m=1}^M (s+1-m)\lambda_m
 =\frac{M+1}{2}N-\sum_{m=1}^M m\lambda_m 
 \label{MaxSpinS}
\end{equation}
can be considered as the maximal total spin, corresponding to the Young diagram $\lambda$. Irreps of $SU(M)$, associated with the Young diagram, can be decomposed into irreps of $R(3)$, having a defined spin. Examples of such decomposition are presented in \cite{kaplan}. Equation \refneq{MaxSpinS} agrees with the maximal spin appearing in these examples. 

The maximal spin \refneq{MaxSpinS} attains its maximum value $N(M-1)/2=N s$ at the one-row Young diagram $\lambda=[N]$. Indeed, for any other Young diagram 
$[N-\sum_{m=2}^M \lambda_m,\lambda_2,\ldots,\lambda_M]$
the maximal spin will be
\begin{equation}
 S=\frac{M-1}{2}N-\sum_{m=2}^M (m-1)\lambda_m \leq \frac{M-1}{2}N
 \label{proofmaxS}
\end{equation}
The one-row Young diagram is associated with a one-dimensional irrep. The basic function is symmetric over all permutations and is an eigenfunction of the total spin.

The normalized character for the conjugate class of transpositions can be expressed as \cite{lassalle2008} 
\begin{eqnarray}
\tilde{\chi}_\lambda (\{2\})=\frac{1}{N(N-1)}
\left[ \sum_{m=1}^M(\lambda_m^2-2 m \lambda_m)+N\right]
\nonumber
\\
=\frac{1}{N(N-1)}
\left( \sum_{m=1}^M\lambda_m^2+2S-M N\right) .
\phantom{qq}
\label{chartran}
\end{eqnarray}
It attains its maximum $\tilde{\chi}_{[N]} (\{2\})=1$ at the one-row Young diagram. Indeed, for any other Young  diagram 
\begin{eqnarray*}
\sum_{m=1}^M\lambda_m^2= N^2
-2\sum_{m=2}^M \lambda_m\left( N-\sum_{m'=2}^M \lambda_{m'}\right) 
\\*
-\sum_{m\neq m'}\lambda_m \lambda_{m'}
\leq N^2 .
\end{eqnarray*}
Taking into account \refeq{proofmaxS}, we get 
$\tilde{\chi}_\lambda (\{2\})\leq 1$.

The zero value of the maximal spin \refneq{MaxSpinS} is reached if $N$ is a multiple of $M$ and $\lambda=[(N/M)^M]$ has $M$ rows of the equal length $N/M$. If $N$ is not a multiple of $M$ ($N=M \lambda_M+k$, $k<M$), the minimum value of the maximal spin $(M-k)k/2$ is attained at the Young diagram $\lambda=[(\lambda_M+1)^k,\lambda_M^{(M-k)}]$. Indeed, for any Young diagram $\lambda'$ the row length can be represented as $\lambda'_m=\lambda_M+1+\Delta\lambda_m$ if 
$m\leq k$ and $\lambda'_m=\lambda_M+\Delta\lambda_m$ if 
$m> k$ with $\sum_{m=1}^M \Delta\lambda_m=0$. The change of the second term in \refeq{MaxSpinS} can be then expressed as
\begin{eqnarray*}
 \sum_{m=1}^M m \Delta\lambda_m=\frac{M+1}{2}\sum_{m=1}^M \Delta\lambda_m
 \nonumber
 \\
 +\sum_{m=1}^{M/2} \left( m-\frac{M+1}{2}\right)  (\Delta\lambda_m-\Delta\lambda_{M+1-m})
\end{eqnarray*}
It is non-positive, since $m-(M+1)/2\leq 0$ and rows of Young diagrams have non-increasing lengths. As a result, we get $S\geq (M-k)k/2$ for any Young diagram.

The first sum in the normalized character \refneq{chartran} can be expressed as
\begin{eqnarray*}
 \sum_{m=1}^M\lambda_m^2=(N+k)\lambda_M+k+2\sum_{m=1}^k\Delta\lambda_m
 +\sum_{m=1}^M\Delta\lambda_m^2
 \nonumber
 \\
 \geq (N+k)\lambda_M+k
\end{eqnarray*}
since $\sum_{m=1}^k\Delta\lambda_m\geq 0$ (otherwise, $\lambda'_m$ will not form a non-increasing sequence). Therefore, 
\[
\tilde{\chi}_\lambda (\{2\})\geq \frac{(N+k)\lambda_M+(M-k+1)k-M N}{N(N-1)} 
\]
and the minimum is attained at 
$\lambda=[(\lambda_M+1)^k,\lambda_M^{M-k}]$.

The boundaries for characters are used for calculation of boundaries for energies.

\section{Selection rules}\label{SI_sel}
Since the total wavefunctions are symmetric (or antisymmetric) over permutations, matrix elements of the $k$-body non-separable interaction $\hat{W}_k(\{j\})$ [see \reflet{\Wk}] are independent of the choice of the interacting particles $j_1,\ldots,j_k$, and, without loss of generality, we can consider the matrix element
\begin{widetext}
 \begin{eqnarray*}
 &&\langle\Psi^{[\lambda']}_{n' l'}|\hat{W}_k(N-k+1,\ldots,N)|\Psi^{[\lambda]}_{n l}\rangle=\frac{N!}{f_\lambda f_{\lambda'}}
 \sum_{t,t',\{N\},\{N'\},r,r'} 
 B^{[\lambda']}_{l' \{N'\} r'}
 B^{[\lambda]}_{l \{N\} r} 
 \nonumber
 \\
 &&
 \times\langle\Phi^{[\lambda']}_{t' n'}|
 \langle\{N'\},\lambda',t',r'|\hat{W}_k(N-k+1,\ldots,N)
 |\{N\},\lambda,t,r\rangle |\Phi^{[\lambda]}_{t n}\rangle .
\end{eqnarray*}
It is expressed, using Eqs.\ (\Psilamnl), \refneq{Xilamtl_Nlamtnu}, \refneq{Nlamtnu}, and \refneq{Blamlnr}, in terms of wavefunctions $|\{N\},\lambda,t,r\rangle$, which do not take into account interactions of spins. These wavefunctions keep the Young diagrams of the total wavefunctions, since the Hamiltonian $\hat{H}_{\mathrm{spin}}$ is diagonal in $\lambda$ (see \refeq{Hspinmatr}). Then the matrix element is expressed in terms of one-body spin states using Eqs.\ \refneq{Nlamtr}, \refneq{XiNP}, and \reflet{\Wk}
\begin{eqnarray}
\langle\{N'\},\lambda',t',r'|\hat{W}_k(N-k+1,\ldots,N)
|\{N\},\lambda,t,r\rangle=\frac{\sqrt{f_\lambda f_{\lambda'}}}{N!}
\sum_{\pr{P},\pr{P}'}D_{r' t'}^{[\lambda']}(\pr{P}')
D_{r t}^{[\lambda]}(\pr{P}) 
\nonumber
\\
\times\prod_{j=1}^{N-k}\delta_{m'_{\pr{P}'j}m_{\pr{P}j}}
\langle m'_{\pr{P}'(N-k+1)}\ldots m'_{\pr{P}'N}|\hat{W}(\mathbf{r}_{N-k+1},\ldots,\mathbf{r}_N)|m_{\pr{P}(N-k+1)}\ldots m_{\pr{P}N}\rangle .
\label{Wk_spin_me}
\end{eqnarray} 
The Kronecker $\delta$-symbols here remain invariant if we replace $\pr{P}'$ by $\pr{P}'\pr{Q}$ and $\pr{P}$ by $\pr{P}\pr{Q}$, where $\pr{Q}$ is any permutation of the first $N-k$ symbols, and the matrix element of 
$\hat{W}(\mathbf{r}_{N-k+1},\ldots,\mathbf{r}_N)$ is independent of these symbols. Therefore, we can average \refeq{Wk_spin_me} over the permutations $\pr{Q}$, replacing the product of the Young matrices by
\[
 \frac{1}{(N-k)!}\sum_{\pr{Q}}D_{r' t'}^{[\lambda']}(\pr{P}'\pr{Q})
D_{r t}^{[\lambda]}(\pr{P}\pr{Q})=\frac{1}{(N-k)!}
\sum_{t'',t'''}D_{r' t'''}^{[\lambda']}(\pr{P}')
D_{r t''}^{[\lambda]}(\pr{P})\sum_{\pr{Q}}
D_{\bar{t}''' \bar{t}'}^{[\bar{\lambda}']}(\pr{Q})
D_{\bar{t}'' \bar{t}}^{[\bar{\lambda}]}(\pr{Q})  ,
\] 
\end{widetext}
where \refeq{ProdYoung} and the reduction to subgroup (see \cite{kaplan}) are used. The Young tableaux $\bar{t}'''$, $\bar{t}'$, $\bar{t}''$, and $\bar{t}$ are obtained by removal of the symbols $N-k+1,\ldots,N$ from the tableaux $t'''$,$t'$, $t''$, and $t$, respectively. Finally, the summation over $\pr{Q}$ using \refeq{OrthYoung} leads to
\[
 \langle\Psi^{[\lambda']}_{n' l'}|\hat{W}_k(N-k+1,\ldots,N)|\Psi^{[\lambda]}_{n l}\rangle\sim \delta_{\bar{\lambda}' \bar{\lambda}} .
\] 
The Young diagrams $\bar{\lambda}'$ and $\bar{\lambda}$ are obtained by removing of $k$ boxes from the diagrams $\lambda'$ and $\lambda$, respectively. Therefore, $\lambda$ and $\lambda'$ can be different by the relocation of no more than $k$ boxes between their rows.

\section{Correlation rules}
Using \reflet{\Psilamnl}, orthogonality of spin wavefunctions, and \refeq{Philamtn}, the expectation value of local spatial correlations \reflet{\hatrhok} can be expressed in terms of wavefunctions $\tilde{\Phi}^{[\lambda]}_{\{\tilde{N}\} t r}$ of non-interacting atoms,
\begin{eqnarray*}
 \langle \Psi^{[\lambda]}_{n l}|\rho_k(\{0\})|\Psi^{[\lambda]}_{n l}\rangle=
 \sum_{\{\tilde{N}\},r,\{\tilde{N'}\},r'}
 \left(\{\tilde{N}\}!\{\tilde{N'}\}!\right) ^{-1/2}
\nonumber
\\
\times
 \sum_{\nu,\nu'}{}' 
A^{[\lambda]}_{n \{\tilde{N}\} \nu}A^{[\lambda]}_{n \{\tilde{N'}\} \nu'}
d_{\nu r}^{[\tilde{\lambda}]}(\{\tilde{N}\})
d_{\nu' r'}^{[\tilde{\lambda}]}(\{\tilde{N'}\})
\nonumber
\\
\times
\frac{1}{f_{\lambda}}\sum_{t}\langle\tilde{\Phi}^{[\lambda]}_{\{\tilde{N'}\} t r'}|\rho_k(\{0\})|\tilde{\Phi}^{[\lambda]}_{\{\tilde{N}\} t r}\rangle  .
\phantom{qqqqqq}
\end{eqnarray*} 
Equation \refneq{tilPhilamNtr} allows us to express the matrix element over $\tilde{\Phi}^{[\lambda]}_{\{\tilde{N}\} t r}$ (the last line of the equation above) in terms of the one-body spatial orbitals $\varphi_{m}(\mathbf{r})$,
\begin{eqnarray}
\frac{1}{N!}\sum_{\pr{P},\pr{P}',t}D_{r' t}^{[\tilde{\lambda}]}(\pr{P}')
D_{r t}^{[\tilde{\lambda}]}(\pr{P})\prod_{j=k+1}^{N}\delta_{m'_{\pr{P}'j}m_{\pr{P}j}}
\nonumber
\\
\times
\int d^{D} r \prod_{j'=1}^k \varphi^*_{m'_{\pr{P'}j'}}(\mathbf{r})
\prod_{j=1}^k \varphi_{m_{\pr{P}j}}(\mathbf{r}).
\label{rhok0tilPhi}
\end{eqnarray} 
The spatial orbitals of the correlating atoms are taken in the same point. Therefore
\[
 \prod_{j'=1}^k \varphi^*_{m'_{\pr{P'}j'}}(\mathbf{r})=\prod_{j'=1}^k \varphi^*_{m'_{\pr{P'}\pr{Q}^{-1}j'}}(\mathbf{r}) \quad (\pr{Q}\in S_k)
\]
and \refeq{rhok0tilPhi} is invariant over permutations $\pr{Q}\in S_k$ of $j'$.
Denoting $\pr{R}=\pr{P'}\pr{Q}^{-1}$, averaging over permutations $\pr{Q}$, and using \refeq{ProdYoung}, \refeq{rhok0tilPhi} can be transformed to
\begin{eqnarray*}
\frac{1}{N!}\sum_{\pr{P},\pr{R},t,t'}D_{r' t'}^{[\tilde{\lambda}]}(\pr{R})
D_{r t}^{[\tilde{\lambda}]}(\pr{P})\prod_{j=k+1}^{N}\delta_{m'_{\pr{R}j}m_{\pr{P}j}}
\\
\times
\int d^{D} r \prod_{j'=1}^k \varphi^*_{m'_{\pr{R}j'}}(\mathbf{r})
\prod_{j=1}^k \varphi_{m_{\pr{P}j}}(\mathbf{r})
\frac{1}{k!}\sum_{\pr{Q}\in S_k} D_{t' t}^{[\tilde{\lambda}]}(\pr{Q}) .
\end{eqnarray*} 
The last sum in this expression can be transformed in the same way as in \refeq{sumP1}
\[
\sum_{\pr{Q}\in S_k} D_{t' t}^{[\tilde{\lambda}]}(\pr{Q})
=k! \delta_{\bar{\tilde{\lambda}} [k]} \delta_{\bar{t'} [0]} \delta_{\bar{t} [0]} ,
\] 
where the Young tableaux $\bar{t'}$ and $\bar{t}$ (obtained by the removal of the symbols $k+1\ldots N$ from the tableaux $t'$ and $t$, respectively) correspond to the same Young diagram $\bar{\tilde{\lambda}}$. The first Kronecker $\delta$ symbol in the above identity zeroes if $\tilde{\lambda}_1<k$. As a result, the expectation values of local spatial correlations vanish if $k>\tilde{\lambda}_1$. Transforming the spatial wavefunctions to the momentum representation, we arrive at the same result for local momentum correlations. 

\section{Multiplet averages of expectation values}\label{SI_MultAv}

Equation (\Psilamnl) and orthogonality of the spin functions lead to the following average of a spin-independent operator $\hat{F}_k$ over a $\lambda$-multiplet,
\begin{eqnarray}
 \bar{F}^{[\lambda]}_k\equiv\frac{1}
 {\tilde{\mathcal{N}}_{\lambda}(N,M_{\mathrm{spat}})}
 \sum_n \langle \Psi^{[\lambda]}_{n l}|\hat{F}_k|\Psi^{[\lambda]}_{n l}\rangle
 \nonumber
 \\
 =\frac{1}{f_{\lambda}
 \tilde{\mathcal{N}}_{\lambda}(N,M_{\mathrm{spat}})}\sum_{t,n}
 \langle\Phi^{[\lambda]}_{t n}|\hat{F}_k|\Phi^{[\lambda]}_{t n}\rangle ,
 \label{Fkexps}               
\end{eqnarray}
where
\[
\tilde{\mathcal{N}}_{\lambda}(N,M_{\mathrm{spat}})=\sum_{\{\tilde{N}\}} \tilde{f}_{\tilde{\lambda}}(\{\tilde{N}\})
\] 
is the total number of the spatial wavefunctions, associated with the Young diagram $\lambda$.
The average \refneq{Fkexps} is independent of the spin quantum numbers $l$.
Since the total wavefunctions are symmetric (or antisymmetric) over permutations, we can suppose, without loss of generality, that the operator $\hat{F}_k$ acts to $\mathbf{r}_1,\ldots,\mathbf{r}_k$. Then equations \refneq{Philamtn}, \refneq{Acomp}, \refneq{tilPhilamNtr}, and \refneq{ProdYoung} lead to
\begin{widetext}
\begin{equation}
 \bar{F}^{[\lambda]}_k=\frac{1}
 { N!  \tilde{\mathcal{N}}_{\lambda}(N,M_{\mathrm{spat}})}
 \sum_{\{\tilde{N}\},r,r'}\frac{1}{\{\tilde{N}\}!}\sum_{\nu}{}' d_{\nu r}^{[\tilde{\lambda}]}
 d_{\nu r'}^{[\tilde{\lambda}]}
 \sum_{\pr{P},\pr{R}}D_{r' r}^{[\tilde{\lambda}]}(\pr{R})
\prod_{j=k+1}^{N}\delta_{m_{\pr{R}\pr{P}j}m_{\pr{P}j}} 
\int d^{Dk} r
 \prod_{j'=1}^k \varphi^*_{m_{\pr{R}\pr{P}j'}}(\mathbf{r}_{j'})\hat{F}_k
 \prod_{j=1}^k  \varphi_{m_{\pr{P}j}}(\mathbf{r}_{j}).
\label{barFkPR}
\end{equation} 
\end{widetext}
The summand in \refeq{barFkPR} is the same
for all $(N-k)!$  permutations $\pr{P}$ corresponding to the given set $\{j\}$ of the symbols 
$j_i=\pr{P}i$ with $1\leq i\leq k$. 
Except of this, Eqs.\ \refneq{Devecval} and \refneq{tildeD} lead to
$\sum'_{\nu} d_{\nu r}^{[\tilde{\lambda}]}
d_{\nu r'}^{[\tilde{\lambda}]}=
\sum_{\nu} d_{\nu r}^{[\tilde{\lambda}]}\delta_{\nu}^{[\tilde{\lambda}]}
d_{\nu r'}^{[\tilde{\lambda}]}/\{\tilde{N}\}!
=\sum_{\pr{P}^{\{\tilde{N}\}}}D_{r r'}^{[\tilde{\lambda}]}(\pr{P}^{\{\tilde{N}\}})/\{\tilde{N}\}!$. Using \refeq{ProdYoung} and taking into account that 
$m_{\pr{P}^{\{\tilde{N}\}}j}=m_j$ for each $j$, 
one gets
\begin{eqnarray*}
 \bar{F}^{[\lambda]}_k=\frac{(N-k)!}
 {N! \tilde{\mathcal{N}}_{\lambda}(N,M_{\mathrm{spat}})}\sum_{\{\tilde{N}\}}\frac{1}{\{\tilde{N}\}!}\sum_{\{j\}}{}'
 \sum_{\pr{R}}\sum_r
 D_{r r}^{[\tilde{\lambda}]}(\pr{R})
 \nonumber
 \\
 \times
\prod_{j'\notin\{j\}}\delta_{m_{\pr{R}j'}m_{j'}}
\int d^{Dk} r
 \prod_{i'=1}^k \varphi^*_{m_{\pr{R}j_{i'}}}(\mathbf{r}_{i'})\hat{F}_k
 \prod_{i=1}^k  \varphi_{m_{j_i}}(\mathbf{r}_{i}),
 \label{barFkQP}
\end{eqnarray*}
where $\sum'_{\{j\}}$ denotes summation over all $j_i$ ($1\leq i\leq k$) such that $j_i\neq j_{i'}$. The summation over $r$ here leads to the trace of the Young matrix --- the character $\chi_{\lambda}$,
\[
 \sum_t D_{t t}^{[\lambda]}(\pr{Q})\equiv 
 \chi_{\lambda}(\pr{Q})=f_{\lambda}\tilde{\chi}_{\lambda}(\pr{Q}) .
\] 
The characters, as well as the normalized characters $\tilde{\chi}_{\lambda}$, are the same for all permutation in a given conjugate class $C_N$ (see \cite{hamermesh,kaplan,pauncz_symmetric}).  
As a result, we get \reflet{\Fkexp} with
\begin{eqnarray}
\langle F_k\rangle_{C_N}=\frac{(N-k)!}
{N!g(C_N)}
\sum_{\{\tilde{N}\}}\frac{1}{\{\tilde{N}\}!}
\sum_{\{j\}}{}'
\sum_{\pr{R}\in C_N}
\prod_{j'\notin\{j\}}\delta_{m_{\pr{R}j'}m_{j'}}
 \nonumber
  \\
 \times
\int d^{Dk} r
 \prod_{i'=1}^k \varphi^*_{m_{\pr{R}j_{i'}}}(\mathbf{r}_{i'})\hat{F}_k
 \prod_{i=1}^k  \varphi_{m_{j_i}}(\mathbf{r}_{i}).
 \phantom{qqqqq}
\label{FkavCk}
\end{eqnarray}

The conjugate classes of the symmetric group $\pr{S}_k$ are characterized by the cyclic structure of the permutations. All permutations in the class 
$C_k=\{k^{\nu_k} \ldots 2^{\nu_2}\}$ have $\nu_l$ cycles of length $l$. This class 
notation omits the number of cycles of the length one, i.e. the number of symbols which
are not affected by the permutations in the class. This number is determined by
the condition $\sum_{l=1}^k l \nu_l=k$. Thus, the same notation can be used for classes $C_k$ and $C_N$
of the groups  $\pr{S}_k$ and  $\pr{S}_N$, respectively.  The 
number of elements in the class $C_N$ of the group $\pr{S}_N$ is expressed as \cite{kaplan,pauncz_symmetric}
\begin{equation}
 g_{C_N}=\frac{N!}{\prod_{l=1}^N \nu_l! l^{\nu_l}}  ,
\label{gclass}
\end{equation}
and the permutations in this class have the parity
\begin{equation}
 \mathrm{sig}(C_N)=\prod_{l=1}^{[N/2]} (-1)^{\nu_{2l}} ,
\end{equation}
where $[x]$ is the integer part of $x$.

If each spatial orbital is occupied only by one particle, $\tilde{N}_m=1$, the set of permutations $\{\pr{P}^{\{\tilde{N}\}}\}$ includes only the identity permutation. In this case, the Kronecker $\delta$-symbols in \refeq{FkavCk} select only the permutations $\pr{R}$ of $k$ symbols $j_i$. All such permutations belong to the conjugate classes $C_k$ of the subgroup $\pr{S}_k$, i.e. $\sum_{l=2}^k l \nu_l\leq k$. Therefore, \reflet{\Fkexp} will include the summation over these classes only. It should be stressed that even in this case,  $\tilde{\chi}_{\lambda}(C_k)$ are characters in irreps of $\pr{S}_N$, i.e. $\sum_{m=1}^M \lambda_m=N$. Equation \refneq{FkavCk} can be simplified in this case by a substitution of $\pr{R}j_{i}=j_{\pr{P}i}$, where $\pr{P}$ are permutations of $k$ symbols,
\begin{eqnarray}
\langle F_k\rangle_{C_k}=\frac{(N-k)!}
{N!g(C_k)}
\sum_{\{j\}}{}'
\sum_{\pr{P}\in C_k}
\int d^{Dk} r
 \prod_{i'=1}^k \varphi^*_{m_{j_{\pr{P}i'}}}(\mathbf{r}_{i'})\hat{F}_k
 \nonumber
  \\*
 \times
 \prod_{i=1}^k  \varphi_{m_{j_i}}(\mathbf{r}_{i}).
 \phantom{qqqqqq}
\label{FkavCk1}
\end{eqnarray}

\section{Multiplet-averaged correlations}\label{SI_corr}
Consider the operator given by \reflet{\hatrhok}. Its expectation values are the $k$-body spatial correlations $\bar{\rho}_k(\{\mathbf{R}\})$. In this case,  \refeq{FkavCk} leads to
\begin{widetext}
\[
\langle \rho_k(\{\mathbf{R}\})\rangle_{C_N}=\frac{(N-k)!}{N!g(C_N)}
\sum_{\{\tilde{N}\}}\frac{1}{\{\tilde{N}\}!}
\sum_{\{j\}}{}'
\sum_{\pr{R}\in C_N}\prod_{j'\notin\{j\}}\delta_{m_{\pr{R}j'}m_{j'}}
\int d^{D}r
\prod_{i=1}^k \varphi^*_{m_{\pr{R}j_{i}}}(\mathbf{r}-\mathbf{R}_{i-1}) \varphi_{m_{j_i}}(\mathbf{r}-\mathbf{R}_{i-1}) 
\]
with $\mathbf{R}_0=0$. 
If each spatial orbital is occupied only by one particle, \refeq{FkavCk1} gives us
\begin{equation}
\langle \rho_k(\{\mathbf{R}\})\rangle_{C_k}=
\frac{(N-k)!}{N!g(C_k)}
\sum_{\{j\}}{}'\sum_{\pr{P}\in C_k}
\int d^{D}r
\prod_{i=1}^k \varphi^*_{m_{j_i}}(\mathbf{r}-\mathbf{R}_{(\pr{P}i)-1}) \varphi_{m_{j_i}}(\mathbf{r}-\mathbf{R}_{i-1}) .
\label{rhokavCk}
\end{equation} 
For the local correlations, with all $\mathbf{R}_{i}=0$, \refeq{rhokavCk} becomes independent of the conjugate class $C_k$ since the integrand therein becomes independent of the permutation $\pr{P}$.

The $k$-body momentum correlations $\bar{g}_k(\{\mathbf{q}\})$ are the expectation values of the operator given by \reflet{\hatgk}. In the coordinate representation, it becomes an integral operator with the kernel,
\[
 (2\pi)^{-D(k-1)}\delta\left( \sum_{j=1}^k(\mathbf{r}'_j-\mathbf{r}_j)\right)
 \exp\left( i \sum_{j=2}^k \mathbf{q}_{j-1}(\mathbf{r}'_j-\mathbf{r}_j)\right) 
\prod_{j=k+1}^N \delta(\mathbf{r}_j-\mathbf{r}'_j) .
\] 
In this case, \refeq{FkavCk} leads to
\begin{eqnarray*}
\langle g_k(\{\mathbf{q}\})\rangle_{C_N}=
\frac{(N-k)!}{(2\pi)^{D(k-1)}N!g(C_N)}&&
\sum_{\{\tilde{N}\}}\frac{1}{\{\tilde{N}\}!}
\sum_{\{j\}}{}'
\sum_{\pr{R}\in C_N}\prod_{j'\notin\{j\}}\delta_{m_{\pr{R}j'}m_{j'}}
\nonumber
\\
&&\times
\int d^{Dk}rd^{Dk}r'
\delta\left( \sum_{i=1}^k(\mathbf{r}'_i-\mathbf{r}_i)\right)
\exp\left( i \sum_{i=2}^k \mathbf{q}_{i-1}(\mathbf{r}'_{i}-\mathbf{r}_i)\right)
\prod_{i=1}^k \varphi^*_{m_{\pr{R}j_{i}}}(\mathbf{r}'_i)\varphi_{m_{j_i}}(\mathbf{r}_i)  .
\end{eqnarray*}
If each spatial orbital is occupied only by one particle, \refeq{FkavCk1} gives us
\begin{equation}
 \langle g_k(\{\mathbf{q}\})\rangle_{C_k}=
\frac{(N-k)!}{(2\pi)^{D(k-1)}N!g(C_k)}
\sum_{\{j\}}{}'\sum_{\pr{P}\in C_k}
\int d^{Dk}r d^{Dk}r'
\delta\left( \sum_{i=1}^k(\mathbf{r}'_i-\mathbf{r}_i)\right)
\exp\left( i \sum_{i=2}^k \mathbf{q}_{i-1}(\mathbf{r}'_{\pr{P}i}-\mathbf{r}_i)\right)
\prod_{i=1}^k \varphi^*_{m_{j_i}}(\mathbf{r}'_i)\varphi_{m_{j_i}}(\mathbf{r}_i)  .
\label{gkavCk}
\end{equation} 
\end{widetext}
This expression becomes independent of the conjugate class $C_k$ if all $\mathbf{q}_j=0$.

\section{Optical lattice}\label{SI_latt}
Consider $N$ cold atoms in a $D$-dimensional optical lattice \cite{bloch2008,*yukalov2009,*svistunov}. Spin-independent interactions between the atoms lead to the Hamiltonian
\begin{equation}
 \hat{H}=\sum_{j=1}^N \left[ -\frac{1}{2} \nabla_j^2 +U_{\mathrm{latt}}(\mathbf{r}_j)+
U(\mathbf{r}_j)\right] +\sum_{j<j'} V(\mathbf{r}_j-\mathbf{r}_{j'}) 
\label{Hlatt}
\end{equation} 
(using units with the mass of the atom and the Plank constant $\hbar$ are equal to $1$)
Here, $\mathbf{r}_j$ is a $D$-dimensional coordinate of the $j$th atom and $\nabla_j$ is the $D$-dimensional gradient. The periodic lattice potential $U_{\mathrm{latt}}(\mathbf{r})$ has $D$ primitive vectors $\mathbf{a}_l$ ($1\leq l\leq D$). The trap potential $U(\mathbf{r})$ is flat on the scale of the lattice period. 

In the case of a deep lattice, the spatial wavefunctions can be expanded in terms of the lowest-band Wannier functions 
$w(\mathbf{r}-\mathbf{T}(\mathbf{n}))$, 
where $\mathbf{T}(\mathbf{n})=\sum_{l=1}^D n_l \mathbf{a}_l$ is the lattice vector, and $\mathbf{n}\equiv(n_1,\ldots,n_D)$ is a $D$-dimensional integer vector. The matrix elements of the terms in $\hat{H}$ will be
\begin{eqnarray}
J_{\mathbf{n}}&=&-\int d^D r w(\mathbf{r}-\mathbf{T}(\mathbf{n}))
\left( -\frac{1}{2} \nabla^2 +U_{\mathrm{latt}}(\mathbf{r})\right)w(\mathbf{r})
\nonumber
\\
V_{\mathbf{n}}&=&\int d^D r d^D r' 
| w(\mathbf{r}'-\mathbf{T}(\mathbf{n}))w(\mathbf{r})|^2 V(\mathbf{r}'-\mathbf{r})
\label{LattMatEl}
\\
U_{\mathbf{n}}&=&U(\mathbf{T}(\mathbf{n}))
\nonumber
\end{eqnarray}
For a zero-range interaction, the exchange interaction will be equal to $V_{\mathbf{n}}$ too. In the unit-filling Mott-insulator regime ($V_0\gg J_{\mathbf{n}}$ for $\mathbf{n}\neq 0$) each lattice site is occupied by one atom. (In experiments, this regime can be realized in the vicinity of the minimum of $U(\mathbf{r})$.) 
Due to the single occupation of each Wannier state, we have $\tilde{D}^{[\lambda]}_{r' r}=\delta_{r' r}$, $\{\tilde{N}\}!=1$, and $d_{\nu r}^{[\lambda]}=\delta_{\nu r}$. Then, using \refneq{tilPhilamNtr} and \refneq{Philamtn} with $\varphi(\mathbf{r})=w(\mathbf{r}-\mathbf{T}(\mathbf{n}))$,
the spatial wavefunction can be expressed as
\begin{equation}
\Phi^{[\lambda]}_{t n}=\sqrt{\frac{f_{\lambda}}{N!}}\sum_{r} 
A^{[\lambda]}_{n r}
\sum_{\pr{P}}D_{t r}^{[\lambda]}(\pr{P}) \mathrm{sig}(\pr{P})
\prod_{j=1}^N w(\mathbf{r}_{\pr{P}j}-\mathbf{T}(\mathbf{n}_j)) ,
\label{Philatt}
\end{equation} 
where $\mathbf{n}_j$ correspond to the occupied sites. 
The coefficients $A^{[\lambda]}_{n r}$ are determined by the eigenproblem  \refneq{HspatAeqEA}, with $\hat{H}_{\mathrm{spat}}=\hat{H}$ of \refeq{Hlatt}. Using orthogonality of the Wannier functions, the matrix elements \refneq{LattMatEl}, and neglecting the hopping 
($J_{\mathbf{n}}$ with $\mathbf{n}\neq 0$), we get the Hamiltonian matrix \refneqs{Hspatmatr}
\[
 \langle \tilde{\Phi}^{[\lambda]}_{\{\tilde{N}\} t r}|\hat{H}|\tilde{\Phi}^{[\lambda]}_{\{\tilde{N}\} t r'}\rangle=
 (-N J_0+\sum_j U_{\mathbf{n}_j})\delta_{r r'}+V^{[\lambda]}_{r r'}
\] 
with
\[
V^{[\lambda]}_{r r'}=\sum_{j<j'}V_{\mathbf{n}_j-\mathbf{n}_{j'}}
[\delta_{r r'}\pm D_{r r'}^{[\lambda]}(\pr{P}_{j j'})]
\] 
The resulting eigenvalue equation for the coefficients $A^{[\lambda]}_{n r}$ and the state energies $E^{[\lambda]}_{n}$ has the form, 
\[
E^{[\lambda]}_{n} A^{[\lambda]}_{n r}=
\sum_{r'}V^{[\lambda]}_{r r'}A^{[\lambda]}_{n r'}  .
\]
The energies are counted from $-N J_0+\sum_j U_{\mathbf{n}_j}$. They form $\lambda$-multiplets with the average energies \reflet{\Eavzero}. In the case of equal lengths of the primitive vectors $\mathbf{a}_i$, taking into account only near-neighbor interactions, 
$V_{\mathrm{near}}=V_{\mathbf{n}}$ with $|\mathbf{n}|=1$, we get
\[
 \bar{E}_{\lambda}=\frac{N N_{\mathrm{near}}V_{\mathrm{near}}}{2}
 \left(  1\pm \frac{N(N-1)}{2}\tilde{\chi}_{\lambda}(\{2\})\right) . 
\]
Here $N_{\mathrm{near}}$ is the number of neighboring sites for each site.

In order to evaluate correlations, let us use the harmonic oscillator approximation for the Wannier functions,
\begin{equation}
 w(\mathbf{r})=\pi^{-D/4}a_{HO}^{-D/2}\exp(-\mathbf{r}^2/(2a_{HO}^2)),
 \label{WanHarmOsc}
\end{equation} 
where $a_{HO}$ is the range of the harmonic potential, approximating the lattice potential in the vicinities of its minima. 

For the spatial correlations, \refeq{rhokavCk} takes the form
\begin{widetext}
\[
\langle \rho_k(\{\mathbf{R}\})\rangle_{C_k}=
\frac{(N-k)!}{N!g(C_k)}
\sum_{\pr{P}\in C_k}\sum_{\{j\}}{}'
\int d^{D}r
\prod_{i=1}^k 
w(\mathbf{r}-\mathbf{T}(\mathbf{n}_{j_i})-\mathbf{R}_{(\pr{P}i)-1}) w(\mathbf{r}-\mathbf{T}(\mathbf{n}_{j_i})-\mathbf{R}_{i-1})  .
\] 
If $\pr{P}=\pr{E}$, the sum over $\{j\}$ can contain terms with 
$\mathbf{T}(\mathbf{n}_{j_i})=-\mathbf{R}_{i-1}$, which are not exponentially small. However, if $C_k\neq \{\}$, even these terms become exponentially small, as $\pr{P}i\neq i$ for at less two $i$, and the integral, calculated with the functions \refneqs{WanHarmOsc}, will be proportional to 
\[
 \exp\left(-\frac{1}{2a_{HO}^2}\left[ \sum_{i=1}^k \left( \Delta \mathbf{R}_i-\frac{1}{k}\sum_{i'=1}^k\Delta \mathbf{R}_{i'}\right)^2
 +\frac{1}{2 k}\left(\sum_{i=1}^k\Delta \mathbf{R}_i\right)^2\right] \right)  .  
\] 
Here $\Delta \mathbf{R}_i=\mathbf{R}_{(\pr{P}i)-1}-\mathbf{R}_{i-1}$.

For the momentum correlations, \refeq{gkavCk} leads to
\begin{eqnarray}
\langle g_k(\{\mathbf{q}\})\rangle_{C_k}=
\frac{(N-k)!}{(2\pi)^{D(k-1)}N!g(C_k)}
\sum_{\pr{P}\in C_k}\sum_{\{j\}}{}'
\exp\left( i\sum_{i=1}^k \mathbf{T}(\mathbf{n}_{j_i})[\mathbf{q}_{i-1}-\mathbf{q}_{(\pr{P}i)-1}]\right)
\nonumber
\\
\times
\int d^{Dk}rd^{Dk}r'
\delta\left( \sum_{i=1}^k(\mathbf{r}'_i-\mathbf{r}_i)\right)\prod_{i=1}^k
\exp\left( i  \mathbf{q}_{i-1}(\mathbf{r}_i-\mathbf{r}'_{i})\right) w(\mathbf{r}'_i)w(\mathbf{r}_i)
\label{gkavCklatts}
\end{eqnarray}
\end{widetext}
with $\mathbf{q}_0=0$. The integral here can be represented as the convolution 
\[
f_k(\{\mathbf{q}\})=\int d^D p \prod_{i=1}^k |\tilde{w}(\mathbf{p}+\mathbf{q}_{i-1})|^2  
\] 
of the Fourier transforms
\[
\tilde{w}(\mathbf{p})=(2\pi)^{-D/2}\int d^Dr w(\mathbf{r}) 
\exp(-i\mathbf{p}\mathbf{r})
\]
of the Wannier function. 

Consider a lattice with the size $L_l$ in the direction of the primitive vector $\mathbf{a}_l$, such that $0\leq n_l < L_l$, and use the equality
\begin{equation}
 \sum_{n_l=0}^{L_l-1} \exp(i \Delta\mathbf{q}_i \mathbf{a}_l n_l)=\frac{\sin (L_l \Delta\mathbf{q}_i \mathbf{a}_l/2)}
 {\sin (\Delta\mathbf{q}_i \mathbf{a}_l/2)} 
 \exp(i \Delta\mathbf{q}_i \mathbf{a}_l \frac{L_l -1}{2}),
 \label{sumexp}
\end{equation} 
were $\Delta\mathbf{q}_i=\mathbf{q}_{i-1}-\mathbf{q}_{(\pr{P}i)-1}$.
Equation \refneq{sumexp} tends to $L_l$ in the limit of $\Delta\mathbf{q}_i\to 0$.
The sum over ${\{j\}}$ in \refeq{gkavCklatts} is the sum over all $j_i$ minus the sums over ${\{j\}}$ where two or more $j_i$ coincide. 
The sum over all $j_i$ is the product of the sums \refneq{sumexp} over all $1\leq l\leq D$ and $1\leq i\leq k$.  
This sum has the maximal value of 
$\prod_{l=1}^D L_l^k=N^k$ when all $\Delta\mathbf{q}_i=0$.
The sum where $k'$ of $j_i$ coincide will have the maximal value of 
$N^{k-k'}$. 
Thus the contributions of such sums can be neglected.
The exponential factors in \refeq{sumexp} will be canceled in all sums over ${\{j\}}$ since 
$
\sum_{i=1}^k\Delta\mathbf{q}_i=\sum_{i=1}^k \mathbf{q}_{i-1}- \sum_{i=1}^k\mathbf{q}_{(\pr{P}i)-1}=0
$. 
As a result, we get 
\begin{eqnarray}
\langle g_k(\{\mathbf{q}\})\rangle_{C_k}&\approx&
\frac{(N-k)!}{N!g(C_k)}f_k(\{\mathbf{q}\})
  \nonumber
  \\
  &&\times
\sum_{\pr{P}\in C_k}\prod_{i=1}^k \prod_{l=1}^D 
\frac{\sin( L_l \mathbf{a}_l\Delta\mathbf{q}_i/2)}{\sin( \mathbf{a}_l\Delta\mathbf{q}_i/2)}  ,
\phantom{qqq}
\label{gkavCklatt}
\end{eqnarray}
oscillating as functions of each component of $\mathbf{q}_j$ with a period $\sim(L_la_l)^{-1}$, except for $C_k=\{\}$, when the arguments of sines in \refeq{gkavCklatt} are equal to $0$. 

If all $\mathbf{q}_{i}=0$, all exponents in \refeq{gkavCklatts} will be equal to zero, and we get \reflet{\bargkzero} with no approximations.

\section{Few-particle cases}
Consider the simplest non-trivial examples of calculation of correlation
functions for few-particle cases. Using \refeq{gclass} for numbers of permutations $g(C_k)$ in the conjugate classes $C_k$ of permutations of $k$ symbols, the universal factors \reflet{\rhotilklam} for lowest-order correlations are expressed as
\begin{eqnarray}
 \tilde{\rho}^{[\lambda]}_2&=&1\pm \tilde{\chi}_\lambda (\{2\})
 \nonumber
 \\
 \tilde{\rho}^{[\lambda]}_3&=&1\pm 3\tilde{\chi}_\lambda (\{2\})
 +2\tilde{\chi}_\lambda (\{3\})
 \\
 \tilde{\rho}^{[\lambda]}_4&=&1\pm 6\tilde{\chi}_\lambda (\{2\})
 +8\tilde{\chi}_\lambda (\{3\})
 \nonumber
 \\
 &&\pm 6\tilde{\chi}_\lambda (\{4\})+3\tilde{\chi}_\lambda (\{2^2\}) .
 \nonumber
\end{eqnarray} 
Here, the sign $+$ or $-$ is taken for bosons or fermions, respectively. The correlations depend on the Young diagrams and are independent of the particle spin whenever the diagram row number does not exceed the multiplicity $M=2s+1$. The particle spin $s$ below means only the minimal spin, allowing the considered Young diagrams, and all results are applicable to particles with higher spins.

For particles with spin $s=1$, the Fortran codes for the normalized characters and the universal factors are presented in the accompanying file \verb|char_corr.f|. They are derived from the explicit expressions for the characters \cite{lassalle2008} and used in Fig. 2. In the simplest non-trivial case of $N=3$ the characters and universal factors are presented in Table \ref{ST_M3N3}.

\begin{table}[!t]
 \caption{The normalized characters and universal factors \reflet{\rhotilklam} for $N=3$ particles of the spin $s=1$ in states associated with the Young diagrams $\lambda$. The upper and lower signs correspond to bosons and fermions, respectively.  \label{ST_M3N3}}
 \begin{ruledtabular}
 \begin{tabular}{c|cc|cc}
 $\lambda $ & $\tilde{\chi}_\lambda (\{2\})$ & $\tilde{\chi}_\lambda (\{3\})$ & $\tilde{\rho}^{[\lambda]}_2$ & $\tilde{\rho}^{[\lambda]}_3$ \\
 \hline
 $[3]$ & 1 & 1 & $1 \pm 1$ & $3 \pm 3$ \\
 $[2 1]$ & 0 & -1/2 & $1$ & $0$ \\
 $[1^3]$ & -1 & 1 & $1 \mp 1$ & $3 \mp 3$ \\
 \end{tabular}
 \end{ruledtabular}
\end{table}

Characters for few particles with higher spins are tabulated in \cite{kaplan}. The characters and universal factors for $N=4$ particles of the spin $s=\frac{3}{2}$ are presented in Table \ref{ST_M4N4}. As would be expected the correlations in Tables \ref{ST_M3N3} and \ref{ST_M4N4} agree with the correlation rules and the correlations for fermions are equal to the correlations for bosons with the conjugate Young diagram.

\begin{table*}
 \caption{The normalized characters and universal factors \reflet{\rhotilklam} for $N=4$ particles of the spin $s=\frac{3}{2}$ in states associated with the Young diagrams $\lambda$. The upper and lower signs correspond to bosons and fermions, respectively.  \label{ST_M4N4}}
 \begin{ruledtabular}
 \begin{tabular}{c|cccc|ccc}
 $\lambda $ & $\tilde{\chi}_\lambda (\{2\})$ & $\tilde{\chi}_\lambda (\{3\})$ & $\tilde{\chi}_\lambda (\{4\})$ & $\tilde{\chi}_\lambda (\{2^2\})$ & $\tilde{\rho}^{[\lambda]}_2$ & $\tilde{\rho}^{[\lambda]}_3$ & $\tilde{\rho}^{[\lambda]}_4$ \\
 \hline
 $[4]$ & 1 & 1 & 1 & 1 & $1 \pm 1$ & $3 \pm 3$ & $12 \pm 12$ \\
 $[3 1]$ & 1/3 & 0 & -1/3 & -1/3 & $1\pm 1/3$ & $1\pm 1$ & $0$ \\
 $[2^2]$ & 0 & -1/2 & 0 & 1 & $1$ & 0 & $0$ \\
 $[2 1^2]$ & -1/3 & 0 & 1/3 & -1/3 & $1\mp 1/3$ & $1\mp 1$ & $0$ \\
 $[1^4]$ & -1 & 1 & -1 & 1 & $1 \mp 1$ & $3 \mp 3$ & $12\mp 12$\\
 \end{tabular}
 \end{ruledtabular}
 \end{table*}
 
As an example of multiple occupation of spatial orbitals consider $N=3$ non-interacting particles of the spin $s=1$ in two spatial orbitals, $\varphi_{1}(\mathbf{r})$ and $\varphi_{2}(\mathbf{r})$, with the occupations $\{\tilde{N_1}\}=2$ and $\{\tilde{N_2}\}=1$, respectively. For non-interacting particles only one term in the sum over $\{\tilde{N}\}$ remains in \refeq{FkavCk}, and for a one-body observable it takes then the form
\[
\langle F_1\rangle_{C_N}=\frac{1}
{N g(C_N)\{\tilde{N}\}!}
\sum_{j=1}^N
\sum_{\pr{R}\in C_N}
\prod_{j'\neq j}\delta_{m_{\pr{R}j'}m_{j'}}
 \langle \varphi_{m_{\pr{R}j}} |\hat{F}_1|\varphi_{m_{j}}\rangle,
\] 
where
\[
 \langle \varphi_{m'} |\hat{F}_1|\varphi_m\rangle = \int d^{D} r
  \varphi^*_{m'}(\mathbf{r})\hat{F}_1
  \varphi_{m}(\mathbf{r}) .
\]
The product of Kronecker symbols above allows only permutations of equal quantum numbers. Then $\pr{R}$ can be either the identity permutation $\pr{E}$ or the transposition $\pr{P}_{12}$, which belong to the conjugate classes $\{\}$ or $\{2\}$, respectively. As a result \reflet{\Fkexp} contains two terms with
\begin{eqnarray}
 \langle F_1\rangle_{\{\}}&=&\frac{1}{6}(2 \langle \varphi_{1} |\hat{F}_1|\varphi_{1}\rangle + \langle \varphi_{2} |\hat{F}_1|\varphi_{2}\rangle) ,
 \phantom{qqq}
 \label{F1iden}
 \\
 \langle F_1\rangle_{\{2\}}&=&\frac{1}{3}\langle F_1\rangle_{\{\}}
 \nonumber
\end{eqnarray}
and the $\lambda$-multiplet-average is expressed as
\[
 \bar{F}^{[\lambda]}_1=f_{\lambda}\langle F_1\rangle_{\{\}}
 (1\pm \tilde{\chi}_\lambda (\{2\}))
\]
with the sign $+$ or $-$ for bosons or fermions, respectively. Finally, using characters in Table \ref{ST_M3N3}, we have
$\bar{F}^{[3]}_1=\bar{F}^{[21]}_1=2\langle F_1\rangle_{\{\}}$, $\bar{F}^{[1^3]}_1=0$ for bosons and $\bar{F}^{[1^3]}_1=\bar{F}^{[21]}_1=2\langle F_1\rangle_{\{\}}$, $\bar{F}^{[3]}_1=0$ for fermions. The results can be applied, for example, to the one-body density matrix $\varrho(\mathbf{r},\mathbf{r}')$ with 
$\langle \varphi_{m} |\varrho(\mathbf{r},\mathbf{r}')|\varphi_m\rangle=\varphi^*_{m}(\mathbf{r})\varphi_{m}(\mathbf{r}')$ in \refeq{F1iden}.

For non-interacting particles and a two-body observable $\hat{F}_2$, \refeq{FkavCk} will be replaced by
\begin{eqnarray*}
\langle F_2\rangle_{C_N}=\frac{2}
{N (N-1) g(C_N)\{\tilde{N}\}!}
\sum_{j_1 < j_2}
\sum_{\pr{R}\in C_N}
\prod_{j_1 \neq j'\neq j_2}\delta_{m_{\pr{R}j'}m_{j'}}
 \nonumber
 \\
 \times
 \langle \varphi_{m_{\pr{R}j_1}} \varphi_{m_{\pr{R}j_2}}|\hat{F}_2|\varphi_{m_{j_1}}\varphi_{m_{j_2}}\rangle,
\end{eqnarray*}
where
\begin{eqnarray*}
 &&\langle \varphi_{m''}\varphi_{m'''} |\hat{F}_2|\varphi_m\varphi_{m'}\rangle = 
 \\
 &&\frac{1}{2}\int d^{D} r_1 d^{D} r_2 
  [\varphi^*_{m''}(\mathbf{r_1})\varphi^*_{m'''}(\mathbf{r_2})\hat{F}_2
  \varphi_{m}(\mathbf{r_1}) \varphi_{m'}(\mathbf{r_1})
 \\ 
  &&+\varphi^*_{m'''}(\mathbf{r_1})\varphi^*_{m''}(\mathbf{r_2})\hat{F}_2
  \varphi_{m'}(\mathbf{r_1}) \varphi_{m}(\mathbf{r_1})].
\end{eqnarray*}
For $N=3$, $\{\tilde{N_1}\}=2$ and $\{\tilde{N_2}\}=1$ the permutations $\pr{R}$, allowed by the Kronecker symbols, depend on $j'$, namely
\begin{eqnarray*}
j'=1 &:& \pr{R}\in \{\pr{E},\pr{P}_{12},\pr{P}_{23},\pr{P}_{123}\}
\\
j'=2 &:& \pr{R}\in \{\pr{E},\pr{P}_{12},\pr{P}_{13},\pr{P}_{132}\}
\\
j'=3 &:& \pr{R}\in \{\pr{E},\pr{P}_{12}\} ,
\end{eqnarray*}
where the cycles of length 3, $\pr{P}_{123}$ and $\pr{P}_{132}$, belong to the conjugate class $\{3\}$. Then  \reflet{\Fkexp} contains three terms with
\begin{eqnarray*}
 \langle F_2\rangle_{\{\}}&=&\frac{1}{6}(\langle \varphi_{1}\varphi_{1} |\hat{F}_2|\varphi_1\varphi_{1}\rangle+2\langle \varphi_{1}\varphi_{2} |\hat{F}_2|\varphi_1\varphi_{2}\rangle)
 \\
 \langle F_2\rangle_{\{2\}}&=&\frac{1}{18}(\langle \varphi_{1}\varphi_{1} |\hat{F}_2|\varphi_1\varphi_{1}\rangle+2\langle \varphi_{1}\varphi_{2} |\hat{F}_2|\varphi_1\varphi_{2}\rangle
 \\
 &&+2\langle \varphi_{2}\varphi_{1} |\hat{F}_2|\varphi_1\varphi_{2}\rangle)
 \nonumber
 \\
 \langle F_2\rangle_{\{3\}}&=&\frac{1}{6}\langle \varphi_{2}\varphi_{1} |\hat{F}_2|\varphi_1\varphi_{2}\rangle
 \nonumber
\end{eqnarray*}
and the $\lambda$-multiplet-average is expressed as
\begin{eqnarray*}
 \bar{F}^{[\lambda]}_2&=&f_{\lambda}\Bigl[\Bigl(\frac{1}{6}\langle \varphi_{1}\varphi_{1} |\hat{F}_2|\varphi_1\varphi_{1}\rangle+\frac{1}{3}\langle \varphi_{1}\varphi_{2} |\hat{F}_2|\varphi_1\varphi_{2}\rangle\Bigr)
 (1\pm \tilde{\chi}_\lambda (\{2\}))
 \\
 &&+\frac{1}{3}\langle \varphi_{2}\varphi_{1} |\hat{F}_2|\varphi_1\varphi_{2}\rangle(\tilde{\chi}_\lambda (\{3\})\pm \tilde{\chi}_\lambda (\{2\}))\Bigr] .
\end{eqnarray*}
Finally, using characters in Table \ref{ST_M4N4}, we have for bosons
\begin{eqnarray*}
 \bar{F}^{[3]}_2&=&\frac{1}{3}(\langle \varphi_{1}\varphi_{1} |\hat{F}_2|\varphi_1\varphi_{1}\rangle+2\langle \varphi_{1}\varphi_{2} |\hat{F}_2|\varphi_1\varphi_{2}\rangle+2\langle \varphi_{2}\varphi_{1} |\hat{F}_2|\varphi_1\varphi_{2}\rangle)
\\
 \bar{F}^{[21]}_2&=&\frac{1}{3}(\langle \varphi_{1}\varphi_{1} |\hat{F}_2|\varphi_1\varphi_{1}\rangle+2\langle \varphi_{1}\varphi_{2} |\hat{F}_2|\varphi_1\varphi_{2}\rangle-\langle \varphi_{2}\varphi_{1} |\hat{F}_2|\varphi_1\varphi_{2}\rangle) .
\\
\bar{F}^{[1^3]}_2&=&0
\end{eqnarray*}
The averages for fermions are equal to the averages for bosons with the conjugate Young diagram. The averages of the local two-body correlations for bosons are expressed as
\begin{eqnarray*}
 \bar{\rho}^{[3]}_2(0)=\frac{1}{3}\left( \int d^{D} r |\varphi_{1}(\mathbf{r})|^4
 +4\int d^{D} r |\varphi_{1}(\mathbf{r})|^2|\varphi_{2}(\mathbf{r})|^2\right)
 \\
 \bar{\rho}^{[21]}_2(0)=\frac{1}{3}\left( \int d^{D} r |\varphi_{1}(\mathbf{r})|^4
 +\int d^{D} r |\varphi_{1}(\mathbf{r})|^2|\varphi_{2}(\mathbf{r})|^2\right) .
\end{eqnarray*}
\end{document}